\documentclass[aps,floatfix,twocolumn,superscriptaddress,amsmath,amssymb,showpacs,10pt]{revtex4-1}
\usepackage{graphicx}
\usepackage{epstopdf}
\usepackage{epsfig}
\usepackage{epsf}
\usepackage{url}
\usepackage[USenglish]{babel}
\usepackage{hyperref}
\def\bcen{\begin{center}}
\def\ecen{\end{center}}
\allowdisplaybreaks
\usepackage{verbatim}
\usepackage{natbib}
\usepackage{amsmath, nccmath}
\usepackage[export]{adjustbox}
\usepackage{dcolumn}
\usepackage{bm}
\usepackage{bbm}
\usepackage{lipsum}
\usepackage{color}
\usepackage{newtxtext}
\usepackage{newtxmath}
\usepackage{sidecap}
\usepackage{tikzit}

\tikzstyle{qtt}=[fill=white, draw=black, shape=rectangle, line width=0.75pt]
\tikzstyle{leftorth}=[fill=white, draw=black, shape=regular polygon, regular polygon sides=3, shape border rotate=270, line width=0.75pt, scale=0.5]
\tikzstyle{rightorth}=[fill=white, draw=black, shape=regular polygon, regular polygon sides=3, shape border rotate=90, line width=0.75pt, scale=0.5]
\tikzstyle{diamond}=[fill=white, draw=black, shape=diamond, line width=0.75pt, scale=0.6]
\tikzstyle{diamondinv}=[fill={black!30}, draw=black, shape=diamond, line width=0.75pt, scale=0.6]
\tikzstyle{blackqtt}=[fill=black, draw=black, shape=rectangle, line width=0.5pt]
\tikzstyle{2-leg tensor}=[fill=white, draw=black, shape=rectangle, minimum width=0.5cm, line width=0.75pt]
\tikzstyle{2-leg tensor round}=[fill=white, draw=black, shape=rectangle, minimum width=0.5cm, line width=0.75pt, rounded corners=3pt, tikzit shape=circle]
\tikzstyle{circle}=[fill=white, draw=black, shape=circle, line width=0.75pt, scale=0.75]
\tikzstyle{blackcircle}=[fill=black, draw=black, shape=circle, line width=0.75pt, scale=0.75]
\tikzstyle{circleinv}=[fill={black!30}, draw=black, shape=circle, line width=0.75pt, scale=0.75]
\tikzstyle{smallcircle}=[fill=white, draw=black, shape=circle, line width=0.75pt, scale=0.5]
\tikzstyle{onehot}=[fill=black, draw=black, shape=circle, line width=0.75pt, scale=0.3]
\tikzstyle{hexagon}=[draw, fill=white, regular polygon, regular polygon sides=6, minimum size=1cm, line width=0.75pt, scale=0.28]
\tikzstyle{pentagon}=[draw, fill=white, regular polygon, regular polygon sides=5, minimum size=1cm, line width=0.75pt, scale=0.28]

\tikzstyle{normal}=[-, line width=0.75pt]
\tikzstyle{dashed}=[-, dashed]

\input{qtt.tikzdefs}

\usepackage[section]{placeins}

\begin{document}
	
	\title{Predictor-corrector method based on dynamic mode decomposition for tensor-train nonequilibrium Green's function calculations}
	
	\author{Maksymilian \'Sroda}
	\affiliation{Department of Physics, University of Fribourg, 1700 Fribourg, Switzerland}
	\author{Ken Inayoshi}
	\affiliation{Department of Physics, Saitama University, Saitama 338-8570, Japan}
	\author{Michael Schüler}
	\affiliation{PSI Center for Scientific Computing, Theory and Data, 5232 Villigen PSI, Switzerland}
	\affiliation{Department of Physics, University of Fribourg, 1700 Fribourg, Switzerland}
	\author{Hiroshi Shinaoka}
        \affiliation{Department of Physics, Saitama University, Saitama 338-8570, Japan}
	\author{Philipp Werner}
	\affiliation{Department of Physics, University of Fribourg, 1700 Fribourg, Switzerland}

\begin{abstract} 
The nonequilibrium Green's function (NEGF) formalism is a powerful tool to study the nonequilibrium dynamics of correlated lattice systems, but its applicability to realistic system sizes and long timescales is limited by unfavorable memory scaling.
While compressed representations, such as the recently introduced quantics tensor train (QTT) format, alleviate the memory bottleneck, the efficiency of QTT-NEGF calculations is hindered by poor initializations and slow or unstable convergence of globally updated self-consistent iterations.
Here, we introduce a predictor-corrector solver for QTT-NEGF simulations that combines dynamic mode decomposition (DMD) extrapolation with the recently proposed causality-preserving block-time-stepping updates. 
The DMD predictor supplies accurate initial guesses that reduce the iteration count of the calculation, while the block-time-stepping correction ensures stable convergence even for long propagation intervals.
Applying this method to the Hubbard model on a $32\times 32$ lattice within the nonequilibrium $GW$ approximation, we demonstrate stable propagation up to times of $t_\mathrm{max}=512$ inverse hoppings, surpassing the capabilities of both matrix-based implementations and previous QTT solvers.
Our contribution is twofold. (i) We integrate tensor dynamic mode decomposition with the QTT representation, which establishes a general framework that is not limited to NEGFs. (ii) We demonstrate its practical benefits in NEGF simulations, where it enables stable and efficient access to unprecedented timescales at high momentum resolution, thereby advancing controlled studies of long-time dynamics and nonequilibrium steady states in correlated lattice systems.
\end{abstract}
	
\maketitle
	
\section{Introduction}

Calculating the nonequilibrium evolution of interacting lattice systems is a theoretical and computational challenge. A successful and versatile approach is the nonequilibrium Green's function (NEGF) technique, where the central object is the momentum ($\mathbf{k}$) dependent two-time Green's function $G_\mathbf{k}(t,t')$. Conventionally, the nonequilibrium Green's function for each $\mathbf{k}$ is stored on the computer as a matrix of values for the discretized time points $(t,t')$. With increasing maximum propagation time $t_\mathrm{max}$, this approach however quickly exhausts the available computer memory. This is because the Kadanoff-Baym equations determining $G_\mathbf{k}(t,t')$ are non-Markovian and involve a memory kernel which depends on the complete history of the system's time evolution. 

To address this challenge, approximation schemes have been developed, including the Generalized Kadanoff Baym Ansatz (GKBA) \cite{Lipavsky1986,Hermanns2012,Kalvova2019,Tuovinen2020,Murakami2020,Schlunzen2020,Joost2020,Karlsson2021,Pavlyukh2022a,Pavlyukh2022b,Pavlyukh2022c,Tuovinen2023,Pavlyukh2024,Bonitz2024}, which reconstructs the two-time Green's function from the time-local density matrix, or memory-truncation schemes \cite{Schueler2018,Stahl2022}, which limit the memory time of the kernel. The former approach has been successfully applied to small nanoclusters \cite{Balzer2013,Bonitz2013,Hermanns2013,Hermanns2014,Joost2022}, but in the case of the state-of-the-art time-linear G1-G2 scheme \cite{Joost2020,Schlunzen2020,Bonitz2024}, its performance in lattice simulations is restricted by the need to store a large momentum-dependent two-particle correlator \cite{Bonitz2024}. The memory truncation scheme, on the other hand, was used in nonequilibrium dynamical mean-field theory calculations \cite{Dasari2021,Ray2025}, where it allowed reaching much longer times than previously accessible. Since this approach involves a control parameter (the truncation time) the accuracy of the simulation can be in principle systematically ensured \cite{Schueler2018,Stahl2022}.

A recent strategy to deal with the memory challenge is to apply memory compression techniques to the nonequilibrium Green's functions, such as the hierarchical off-diagonal low-rank \cite{Kaye2021} or quantics tensor-train (QTT) decompositions \cite{MSSTA,Murray2024,Sroda2024}. 
By encoding the information contained in the functions in a more efficient way, the memory demand can be significantly reduced without compromising the accuracy of the simulation. The challenge, however, becomes the integration of these compression techniques into the standard procedures for solving diagrammatic many-body problems.   
From this point of view, the QTT representation is particularly promising as it leverages the established tensor-network toolbox to efficiently perform standard linear-algebra operations in the compressed form.

If one a priori knows the exact lattice self-energy $\Sigma_\mathbf{k}(t,t')$, then solving the Dyson equation once is enough to obtain $G_\mathbf{k}(t,t')$ and hence to solve the nonequilibrium problem. However, in standard weak-coupling diagrammatic schemes, the self-energy is given in terms of the Feynman diagrams as a functional of the Green's function, $\Sigma_\mathbf{k}=\Sigma_\mathbf{k}[G_\mathbf{k}]$. One thus needs to solve the equations determining $\Sigma_\mathbf{k}$ and $G_\mathbf{k}$ self-consistently. The convergence properties of such self-consistent iterations are strongly dependent on the initial starting point. The common choice is the noninteracting Green's function $G_{0\mathbf{k}}(t,t')$, which can be calculated explicitly. Except in the very weakly correlated regime, this function however differs significantly from the interacting solution, which can lead to slow convergence or even to numerical instabilities in the self-consistency loop. This is particularly the case for global updates of $G_\mathbf{k}(t,t')$, where one solves the equations of motion for all $(t,t')$ points simultaneously \cite{Sroda2024,Gasperlin2025,Inayoshi2025}, as opposed to the conventionally applied time stepping \cite{Nessi,stefanucci_van_leeuwen_2013}. Such global update schemes are the simplest to integrate with compression schemes \cite{Murray2024,Sroda2024}. To mitigate this problem, a causality-based divide-and-conquer solver was recently introduced \cite{Inayoshi2025} to improve the stability and performance of QTT-based nonequilibrium dynamical mean-field theory simulations.
While this new solver enabled iterative and stable extensions of the time domain, the initial guesses on the extensions were taken to be the noninteracting Green's function, which is suboptimal in most applications, slowing down convergence.

In this paper, we develop a predictor-corrector solver to further improve the quantics-tensor-train nonequilibrium Green's function methodology.
The prediction step is based on dynamic mode decomposition (DMD) \cite{Schmid2008,Schmid2010,Tu2014,LeClainche2017,Schmid2022,Baddoo2023}, a data-driven approach that approximates nonlinear dynamics by fitting a low-rank linear model to short-time trajectories of system observables. 
The resulting spatio-temporal modes not only give insight into the system's underlying dynamical structure \cite{Schmid2010b,Sayadi2014,Tu2014,Kutz2016,Schmid2022,Baddoo2023} but also allow reliable extrapolation, extending the dynamics up to twice the length of the known data or more \cite{Yin2022,Yin2023,Reeves2023,Maliyov2024,Kaneko2025}.
This gives the DMD a wide range of applications in diverse fields, for instance, in fluid dynamics \cite{Schmid2010b,Sayadi2014,Tu2014,Schmid2022}, infectious disease modeling \cite{Proctor2015}, neuroscience \cite{McLean2024}, climate modeling \cite{Mankovich2025}, metabolic dynamics \cite{Wustner2025}, materials science \cite{Simha2024}, and nonequilibrium Green's function simulations \cite{Yin2022,Yin2023,Reeves2023}, to name just a few.
Crucially, DMD can be performed directly in a tensor representation \cite{Klus2018,Li2023,Yin2025}, which makes it particularly attractive for our purposes.

Here, we integrate DMD with the \emph{quantics} tensor-train representation, which, to our knowledge, has not been considered before.
This integration enables DMD to be carried out efficiently at exponentially fine time resolution and may also provide a promising route to denoise QTTs obtained from simulations with finite numerical accuracy.
We present the algorithm in a general way in order to facilitate its future applications beyond NEGF calculations.
The correction step proceeds according to a \emph{block time stepping} strategy \cite{Inayoshi2025}. We solve the Dyson equation on successive block time steps which preserve the very fine quantics resolution and are orders of magnitude larger than the individual time steps in matrix-based NEGF implementations. The combination of a (quasi)causal solver with a good starting guess leads to a stable, predictable and accelerated convergence at each block time step.
The method is tested by considering interaction quenches in the time-dependent Hubbard model on a $32 \times 32$ square lattice within the $GW$ approximation. We show that the new scheme surpasses both our previous implementation~\cite{Sroda2024} and the capabilities of conventional matrix-based methods, reaching propagation times as large as $t_\mathrm{max}=512$.
Our approach defines the current state of the art in NEGF calculations of large translationally invariant systems with full memory kernel.

The paper is organized as follows.
In Sec.~\ref{formalism}, we review the basics of nonequilibrium Green's function theory (Sec.~\ref{negf}) and the quantics tensor train representation (Sec.~\ref{qtt}).
This sets the stage for Sec.~\ref{method}, which details the quantics-tensor-train dynamic mode decomposition (Sec.~\ref{dmd}) and the predictor-corrector solver (Sec.~\ref{predictor-corrector}).
Finally, Sec.~\ref{results} presents numerical results demonstrating our approach on the example of nonequilibrium dynamics in the square-lattice Hubbard model.
The paper concludes with Sec.~\ref{conclusions}.

\section{Theoretical and numerical formalism}\label{formalism}

\subsection{Nonequilibrium Green's functions}\label{negf}

Consider a system with the generic Hamiltonian
\begin{equation}
	H(t) = \sum_{\mathbf{k}\sigma} \epsilon_\mathbf{k\sigma}(t) c^\dagger_{\mathbf{k}\sigma}c_{\mathbf{k}\sigma} + H_\mathrm{int}(t),\label{eq:genericH}
\end{equation}
where $c^\dagger_{\mathbf{k}\sigma}$ creates a fermion with momentum $\mathbf{k}$ and spin $\sigma$, $\epsilon_{\mathbf{k}\sigma}(t)$ is the time-dependent electron dispersion with the related hopping amplitude $t_\mathrm{h}$, and $H_\mathrm{int}(t)$ is the time-dependent interaction term of the Hamiltonian. Both $\epsilon_{\mathbf{k}\sigma}(t)$ and $H_\mathrm{int}(t)$ will be given below for specific applications. 
Energy is expressed in units of hoppings $t_\mathrm{h}$ and time is measured in units of inverse hoppings $1/t_\mathrm{h}$ ($\hbar \equiv 1$).
We suppress the spin index in the following discussion, since we will focus on paramagnetic states.

The nonequilibrium dynamics of the system~\eqref{eq:genericH} is encoded in the interacting contour-ordered Green's function
\begin{equation}
	G_{\mathbf{k}}(t,t') = -i \langle \mathcal{T}_\mathcal{C} c_{\mathbf{k}}(t) c^\dagger_{\mathbf{k}}(t') \rangle.\label{eq:negf}
\end{equation}
Here, $\mathcal{T}_\mathcal{C}$ is the time-ordering operator on the Kadanoff-Baym contour, \includegraphics[height=.7em]{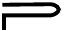}, which runs along $0 \to t_\text{max} \to 0 \to -i\beta$ in the complex time plane, with $t_\text{max}$ the maximum time and $\beta$ the inverse temperature. For details on the nonequilibrium Green's function formalism we refer, e.g., to Ref.~\cite{Aoki2014}, while here we only discuss the essential aspects.

To obtain the above Green's function, one needs to solve the nonequilibrium Dyson equation
\begin{equation}
	G_\mathbf{k}(t,t') = G_{0\mathbf{k}}(t,t') + [G_{0\mathbf{k}}*\Sigma_\mathbf{k}*G_\mathbf{k}](t,t'),\label{eq:dyson}
\end{equation}
where $G_{0\mathbf{k}}(t,t') = -i[\theta_\mathcal{C}(t,t')-f_\beta(\epsilon_\mathbf{k}(0))]e^{-i\int_{t'}^t d\bar{t}\epsilon_\mathbf{k}(\bar{t})}$ is the noninteracting Green's function \cite{Freericks2005} with $f_\beta(\omega)$ the Fermi function and $\theta_\mathcal{C}(t,t')$ the contour step function \cite{Aoki2014}, $[a*b](t,t') = \int_\mathcal{C} d\bar{t} a(t,\bar{t}) b(\bar{t},t')$ denotes a contour convolution, and $\Sigma_\mathbf{k}(t,t')$ is the self-energy describing many-body effects due to $H_\mathrm{int}$. 
In the diagrammatic approaches considered here, the self-energy can be expressed as a functional of $G_\mathbf{k}$, $\Sigma_\mathbf{k}=\Sigma_\mathbf{k}[G_\mathbf{k}]$, in terms of a chosen set of Feynman diagrams.
When solving the Dyson equation in the integral form, it is convenient to view it as a linear problem,
\begin{equation}
	(1-G_{0\mathbf{k}}*\Sigma_\mathbf{k}*{}\!) \, G_\mathbf{k} = G_{0\mathbf{k}},\label{eq:dysonlinear}
\end{equation}
which can be further decomposed into the so-called Kadanoff-Baym equations (KBE). These are four coupled integral equations for the ``physical'' Green's functions, $G_\mathbf{k}^M$ (Matsubara), $G_\mathbf{k}^R$ (retarded), $G_\mathbf{k}^\rceil$ (left-mixing) and $G_\mathbf{k}^<$ (lesser), which have the time arguments restricted to different branches of the contour. Using the Langreth rules \cite{Langreth1976,stefanucci_van_leeuwen_2013}, one can write the equations explicitly as
\begin{align}
    &(1-G_0^M\star\Sigma^M\star{})\,G^M = G_0^M,\label{eq:mat}\\[.3em]
    &(1-G_0^R\cdot\Sigma^R\cdot{})\,G^R = G_0^R,\label{eq:ret}\\[.3em]
    &\begin{aligned}
        (1-G_0^R\cdot\Sigma^R\cdot{})&\,G^\rceil = G_0^\rceil\\
		&+ (G_0^R \cdot \Sigma^\rceil + G_0^\rceil \star \Sigma^M) \star G^M,
    \end{aligned}\label{eq:tv}\\[.3em]
    &\begin{aligned}
        (1-G_0^R\cdot\Sigma^R\cdot{})&\,G^< = G_0^<\\
		&+ (G_0^R \cdot \Sigma^<+G_0^\rceil \star \Sigma^\lceil + G_0^< \cdot \Sigma^A) \cdot G^A\\
        {}&+ (G_0^R \cdot \Sigma^\rceil + G_0^\rceil \star \Sigma^M )\star G^\lceil ,
    \end{aligned}\label{eq:les}
\end{align}
where $[a\cdot b](t,t') = \int_0^{t_\mathrm{max}} d\bar{t} a(t,\bar{t}) b(\bar{t},t')$ and $[a\star b](t,t') = -i\int_0^{\beta} d\bar{t} a(t,\bar{t}) b(\bar{t},t')$ are convolutions on the real- and imaginary-time axes, respectively, while the $\mathbf{k}$ index was suppressed for clarity. In the above, also the advanced ($A$) and right-mixing ($\lceil$) components appear, which can be obtained from the other four by symmetry operations \cite{Aoki2014}. Note that the Matsubara Green's function is defined here with an extra $i$ factor, as compared to the usual convention. That is, $G^M(t,t') = -i\langle \mathcal{T}_M c(t)c^\dagger(t')\rangle$, with both arguments on the imaginary-time axis and with $\mathcal{T}_M$ the corresponding time-ordering operator.

Because $\Sigma_\mathbf{k}=\Sigma_\mathbf{k}[G_\mathbf{k}]$, Eqs.~\eqref{eq:mat}-\eqref{eq:les} define a self-consistent loop.
After choosing an initial ansatz for $G_\mathbf{k}$ (usually $G_{0\mathbf{k}}$), one uses it to obtain the self-energy, which is then used to calculate an updated $G_\mathbf{k}$, and the process is repeated until convergence.
Here, we perform a direct fixed-point iteration and do not apply any mixing between the iterations, as it is not necessary for stabilization owing to the reliable initial guess from DMD.
Since the Matsubara equation \eqref{eq:mat} defines the equilibrium initial state at $t<0$, it is independent of the other components and the Matsubara Green's function can be calculated self-consistently as a first step. The remaining equations are then solved one by one in the order \eqref{eq:ret}-\eqref{eq:les} at each iteration of the self-consistency loop.

In this work, we present a procedure to obtain a better ansatz than $G_{0\mathbf{k}}$ in cases where one already has a solution on a short contour and wants to use this information to produce a good starting point for the self-consistent iterations on a longer contour.
This is particularly important when globally updating the Kadanoff-Baym equations, i.e., updating all $(t,t')$ points simultaneously, without invoking causality to keep the solution on past times fixed. In such a case, the iterative calculation has a particular tendency to become slow or unstable except on rather short contours \cite{Sroda2024}.
The procedure also helps to speed up convergence in a time-stepping calculation with potentially large time steps.

In the numerical implementation, Eqs.~\eqref{eq:mat}-\eqref{eq:les} are discretized. We choose an equidistant grid with spacing $dt$ on the real- and $d\tau$ on the imaginary-time axis. The convolutions thus become matrix multiplications, $[a\cdot b]_{t,t'} = \sum_{\bar{t}} a_{t,\bar{t}} w_{\bar{t}} b_{\bar{t},t'}$, $[a \star b]_{t,t'} = \sum_{\bar{t}} a_{t,\bar{t}} v_{\bar{t}} b_{\bar{t},t'}$, where $w$, $v$ are diagonal matrices containing the integration weights for the real- and imaginary-time axes, respectively. Here, we adopt the trapezoidal rule $w=dt\,\mathrm{diag}(\frac{1}{2},1,\ldots,1,\frac{1}{2})$, $v=-id\tau\,\mathrm{diag}(\frac{1}{2},1,\ldots,1,\frac{1}{2})$, $\theta_\mathcal{C}(t,t) = 1/2$, which should lead to negligible discretization errors for the very fine grids ($dt, d\tau \lesssim 10^{-6}$) that we use (see below).

\subsection{Quantics-tensor-train compression of nonequilibrium Green's functions}\label{qtt}
To mitigate the memory demand for storing and manipulating the nonequilibrium Green's function $G_{\mathbf{k}}(t,t')$, we represent it as a quantics tensor train. Concretely, for each $\mathbf{k}$ point we store four tensor trains of the form 
\begin{equation}
	G_\mathbf{k}^c(t,t') = \begin{tikzpicture}
	\begin{pgfonlayer}{nodelayer}
		\node [style=qtt] (0) at (-3, 0.125) {};
		\node [style=qtt] (1) at (-2, 0.125) {};
		\node [style=qtt] (2) at (-1, 0.125) {};
		\node [style=qtt] (3) at (0, 0.125) {};
		\node [style=none] (4) at (1, 0.125) {$\ldots$};
		\node [style=qtt] (5) at (2, 0.125) {};
		\node [style=qtt] (6) at (3, 0.125) {};
		\node [style=none, label={below:$t_1\vphantom{'}$}] (7) at (-3, -0.4) {};
		\node [style=none, label={below:$t'_1$}] (8) at (-2, -0.4) {};
		\node [style=none, label={below:$t_2\vphantom{'}$}] (9) at (-1, -0.4) {};
		\node [style=none, label={below:$t'_2$}] (10) at (0, -0.4) {};
		\node [style=none, label={below:$t_R\vphantom{'}$}] (12) at (2, -0.4) {};
		\node [style=none, label={below:$t'_R$}] (13) at (3, -0.4) {};
		\node [style=none] (14) at (0.5, 0.125) {};
		\node [style=none] (15) at (1.5, 0.125) {};
	\end{pgfonlayer}
	\begin{pgfonlayer}{edgelayer}
		\draw [style=normal] (0) to (7.center);
		\draw [style=normal] (1) to (8.center);
		\draw [style=normal] (2) to (9.center);
		\draw [style=normal] (3) to (10.center);
		\draw [style=normal] (5) to (12.center);
		\draw [style=normal] (6) to (13.center);
		\draw [style=normal] (6) to (5);
		\draw [style=normal] (3) to (2);
		\draw [style=normal] (2) to (1);
		\draw [style=normal] (1) to (0);
		\draw [style=normal] (3) to (14.center);
		\draw [style=normal] (5) to (15.center);
	\end{pgfonlayer}
\end{tikzpicture}\label{eq:negfqtt},
\end{equation}
where $c \in \{M,R,\rceil,<\}$ distinguishes the four Green's function components and the standard tensor graphical notation is assumed. A tensor train is drawn as a chain of sites, with ``leg'' indices corresponding to the function's arguments, connected by bonds representing sums (contractions) over repeated auxiliary indices.
The format~\eqref{eq:negfqtt} corresponds to a factorization of the conventional matrix representation into a product of three-way tensors, $G_\mathbf{k}^c(t,t') = \begin{tikzpicture}
	\begin{pgfonlayer}{nodelayer}
		\node [style=2-leg tensor] (3) at (0, 0.125) {};
		\node [style=none] (10) at (-0.25, -0.4) {};
		\node [style=none] (11) at (0.25, -0.4) {};
		\node [style=none] (12) at (-0.25, 0) {};
		\node [style=none] (13) at (0.25, 0) {};
	\end{pgfonlayer}
	\begin{pgfonlayer}{edgelayer}
		\draw [style=normal] (10.center) to (12.center);
		\draw [style=normal] (11.center) to (13.center);
	\end{pgfonlayer}
\end{tikzpicture} = \begin{tikzpicture}
	\begin{pgfonlayer}{nodelayer}
		\node [style=2-leg tensor, minimum width=.9cm] (3) at (0, 0.125) {};
		\node [style=none] (11) at (0.6, -0.4) {};
		\node [style=none] (13) at (0.6, 0) {};
		\node [style=none] (14) at (0.15, -0.3) {...};
		\node [style=none] (15) at (0.75, -0.4) {};
		\node [style=none] (16) at (0.75, 0) {};
		\node [style=none] (19) at (-0.45, -0.4) {};
		\node [style=none] (20) at (-0.45, 0) {};
		\node [style=none] (21) at (-0.3, -0.4) {};
		\node [style=none] (22) at (-0.3, 0) {};
		\node [style=none] (23) at (-0.6, -0.4) {};
		\node [style=none] (24) at (-0.6, 0) {};
		\node [style=none] (25) at (-0.75, -0.4) {};
		\node [style=none] (26) at (-0.75, 0) {};
	\end{pgfonlayer}
	\begin{pgfonlayer}{edgelayer}
		\draw [style=normal] (11.center) to (13.center);
		\draw [style=normal] (15.center) to (16.center);
		\draw [style=normal] (19.center) to (20.center);
		\draw [style=normal] (21.center) to (22.center);
		\draw [style=normal] (23.center) to (24.center);
		\draw [style=normal] (25.center) to (26.center);
	\end{pgfonlayer}
\end{tikzpicture} \approx \begin{tikzpicture}
	\begin{pgfonlayer}{nodelayer}
		\node [style=qtt] (0) at (-2, 0.125) {};
		\node [style=qtt] (1) at (-1, 0.125) {};
		\node [style=none] (4) at (0, 0.125) {$\ldots$};
		\node [style=qtt] (5) at (1, 0.125) {};
		\node [style=qtt] (6) at (2, 0.125) {};
		\node [style=none] (7) at (-2, -0.4) {};
		\node [style=none] (8) at (-1, -0.4) {};
		\node [style=none] (12) at (1, -0.4) {};
		\node [style=none] (13) at (2, -0.4) {};
		\node [style=none] (14) at (-0.5, 0.125) {};
		\node [style=none] (15) at (0.5, 0.125) {};
	\end{pgfonlayer}
	\begin{pgfonlayer}{edgelayer}
		\draw [style=normal] (0) to (7.center);
		\draw [style=normal] (1) to (8.center);
		\draw [style=normal] (5) to (12.center);
		\draw [style=normal] (6) to (13.center);
		\draw [style=normal] (6) to (5);
		\draw [style=normal] (1) to (0);
		\draw [style=normal] (5) to (15.center);
		\draw [style=normal] (1) to (14.center);
	\end{pgfonlayer}
\end{tikzpicture}$\,, and one can keep the mental picture of a matrix when manipulating the train. 
The multiple leg indices follow from enumerating the time points $t$, $t'$ by binary numbers $[t_1,\ldots,t_R]_2, [t'_1,\ldots,t'_R]_2 \in \{{0,1,\ldots,2^R-1}\}$ with $t_n, t'_n \in \{0,1\}$ the binary digits. 
Taking $t,t' \in [0,t_\mathrm{max}]$, we explicitly have
\begin{equation}
	t = \frac{t_\mathrm{max}}{2^R-1}[t_1,\ldots,t_R]_2 = \frac{t_\mathrm{max}}{2^R-1}\sum_{n=1}^R 2^{R-n}\, t_n,\label{eq:binary}
\end{equation}
and analogously for $t'$. This corresponds to the discretization of the two-time plane $[0,t_\mathrm{max}] \times [0,t_\mathrm{max}]$ into a $2^R \times 2^R$ grid with spacing $dt=t_\mathrm{max}/(2^R-1)$, where, in typical applications, $R \sim 30$. It is implicitly understood that whenever $t$ or $t'$ lie on the imaginary-time axis, $t_\mathrm{max}\equiv \beta$, and that in general the number of bits $R$ for the imaginary- and real-time axes can be different (corresponding to unequal $dt$, $d\tau$ and/or $t_\mathrm{max}$, $\beta$). Note that all other two- and one-time functions needed in the calculation are also stored in the form \eqref{eq:negfqtt}.

Although the binary encoding \eqref{eq:binary} is exact, the factorization \eqref{eq:negfqtt} is approximate. However, its accuracy is fully controllable by adjusting the dimensions, or ranks, $D$ of the auxiliary bond indices of the tensor train. This is done either by setting a fixed maximum bond dimension $D_\mathrm{max}$ or by demanding that the squared Frobenius-norm error at each
tensor of the train is below a chosen tolerance, $|T-\tilde{T}|^2_\mathrm{F}/|T|^2_\mathrm{F} < \tau_\mathrm{cutoff}$, where $T$, $\tilde{T}$ denote the original and truncated tensors, respectively.
We choose to define the tolerance in terms of the squared Frobenius norm following the convention in the ITensor library \cite{itensor}.

The strength of the quantics representation is that each tensor describes a distinct length scale on which the function varies: from coarse scales on the left to fine scales on the right. Furthermore, these length scales are exponential. Extending the length $R$ of the binary string
by one adds only two tensors to the train but doubles the size
of the underlying matrix along both dimensions, either doubling the domain or halving the grid spacing. Crucially, factorizing functions in terms of length scales generically results in low bond dimensions, $D_\mathrm{max} \sim \mathcal{O}(100)$, for many functions encountered in physics, in particular the many-body correlation functions \cite{MSSTA}. This leads to very high compression ratios, as a QTT stores only $\mathcal{O}(D^2_\mathrm{max}\times\mathrm{length})$ elements. For example, the lesser equilibrium noninteracting Green's function in the Hamiltonian's eigenbasis has $D=1$, as it is a product of two exponentials $e^{-i \epsilon_\mathbf{k}t} \, e^{i \epsilon_\mathbf{k}t'}$, which factorize perfectly using Eq.~\eqref{eq:binary}.

The empirical observation of low-rankness of physical functions is related to the concept of length-scale separation. Low $D$ means that the individual scales are only weakly ``entangled'' or correlated with one another, with correlations mainly between the neighboring scales [hence the bits $t_n$, $t'_n$ corresponding to the same scale are placed as neighbors in Eq.~\eqref{eq:negfqtt}]. This intuition can be put on more rigorous grounds by viewing a QTT as a multiscale polynomial interpolation \cite{Lindsey2024}, beyond, e.g., an interpretation in terms of Fourier analysis \cite{Dolgov2012}. For instance, this point of view allows one to understand why functions with sharp features can still have low-rank QTT representations, e.g., the delta function has $D=1$. It can also be used to establish bounds on the bond dimensions, and to explain why the bond dimensions decay in the direction of the fine scales, eventually reaching $D=1$ once all the fine features are resolved. It is the latter property that allows us to use exponentially fine grids at minimal extra cost, enabling, e.g., accurate integration with low-order quadratures.

Importantly, casting all objects into the tensor train format gives access to the wide toolbox of efficient linear-algebra algorithms devised by the tensor-network community.
All operations needed to solve the NEGF problem of Sec.~\ref{negf} can be performed without ever leaving the QTT form \cite{Murray2024,Sroda2024}.
The input preparation is done directly [$\mathcal{O}(1)$] or via tensor cross interpolation \cite{Oseledets2011,Dolgov2020,Fernandez2022,Ritter2024,Fernandez2024} [$\mathcal{O}(D^3)$], matrix multiplications (convolutions) and element-wise multiplications (needed to calculate diagrams) correspond to tensor network contractions [$\mathcal{O}(D^4)$], summations are simply TT sums [$\mathcal{O}(D^3)$], while the solution of the linear problem in Eqs.~\eqref{eq:mat}-\eqref{eq:les} is done with an algorithm similar to density-matrix renormalization group (DMRG) that locally updates the train [$\mathcal{O}(D^4)$]; see Ref.~\cite{Sroda2024}. 
For a further discussion of the quantics representation in various applications to many-body correlation functions, including the description of basic linear-algebra operations, see also Ref.~\cite{MSSTA}. 
Here, we focus on supplementing the solution procedure of Ref.~\cite{Sroda2024} with a DMD extrapolated guess and a quasicausal predictor-corrector method.

\section{Method}\label{method}

\subsection{Dynamic mode decomposition with quantics tensor trains}\label{dmd}

While matrix-based dynamic mode decomposition (DMD) \cite{Schmid2008,Schmid2010,Tu2014,LeClainche2017,Schmid2022,Baddoo2023} was previously applied to predict dynamics encoded in NEGFs \cite{Yin2022,Yin2023,Reeves2023}, it was not used in the context of a predictor-corrector method. Instead, the idea was to avoid the cost of solving the KBE for large $t_\mathrm{max}$ and to infer the long-time behavior of the system from extrapolated data. This approach can be successfully applied to systems with damped oscillatory behavior, but it is hard to judge the accuracy of the extrapolated dynamics in systems exhibiting prethermalization or other complex physical phenomena. 

Below, we provide a general exposition on how to apply DMD to fit and extrapolate arbitrary functions stored in the QTT format \eqref{eq:negfqtt}. A similar tensor-based DMD algorithm was considered before in a different setting \cite{Klus2018,Li2023,Yin2025} but, to our knowledge, has not been combined with the quantics representation. In our applications, we will use the DMD predictions only to produce an initial guess for the self-consistent iteration of the KBE. 

The premise of DMD is to approximate the dynamics of a nonlinear dynamical system by finding its best low-rank linearization. This is done in a data-driven fashion by fitting the linear model to the known dynamics of the system's ``observables'' within a short time window. 
The DMD requires only that the known data are connected via some unknown dynamical system,  
irrespective of the data's actual physical interpretation \cite{Tu2014,Yin2022}.
In fact, smoothness between different observables at a fixed time $t$ is not necessary, nor does the time sampling need to be sequential \cite{Tu2014}, enabling a very diverse set of quantities to be used in constructing the model.

Let us briefly recall the standard procedure.
Assume that we know the dynamics of several physical quantities, or observable functions, $x_i(t)$, arranged into the data matrix $X=\bigl(\, x_i(0)\; x_i(dt)\; \cdots\; x_i(m\,dt)\; \cdots\; x_i(t_\mathrm{max})\,\bigr)$. Its columns $m$ correspond to consecutive time points and rows $i$ to the different observables, such that each column contains all observables measured at fixed $t$.
First, $X$ is split into two matrices 
\begin{equation}
	\begin{aligned}
		X_1 &= 
		\begin{pmatrix}
			\vdots & \vdots &        & \vdots          &        & \vdots                  \\
			x_i(0) & x_i(dt) & \cdots & x_i(m\,dt) & \cdots & x_i(t_{\text{max}}-dt) \\
			\vdots & \vdots &        & \vdots          &        & \vdots                 
		  \end{pmatrix},\\
		X_2 &= 
		\begin{pmatrix}
			\vdots & \vdots &        & \vdots          &        & \vdots                  \\
			x_i(dt) & x_i(2dt) & \cdots & x_i\big((m+1)\,dt\big) & \cdots & x_i(t_{\text{max}})  \\
			\vdots & \vdots &        & \vdots          &        & \vdots             
		  \end{pmatrix},
	\end{aligned}\label{eq:X1X2matrices}
\end{equation}
where, importantly, the data in $X_2$ are shifted one time step forward with respect to the data in $X_1$.
One then assumes a linear evolution $A X_1 = X_2$ over a single time step according to the operator $A$. The least-squares solution is $A = X_2 X_1^{-1}$, with $X_1^{-1}$ understood as the Moore-Penrose pseudoinverse. In practice, one computes a truncated singular value decomposition (SVD) $X_1 = U S V^\dagger$ to find $X_1^{-1} = V S^{-1} U^\dagger$, which is the main computational cost of DMD. Projecting onto the dominant left singular vectors gives the reduced operator $\tilde{A} = U^\dagger A U = U^\dagger X_2 V S^{-1}$, whose eigendecomposition $\tilde{A} = W \Lambda W^{-1}$ yields reduced eigenvalues and eigenvectors. 
The full dynamic modes, i.e., the eigenvectors of $A$, are recovered as $\Phi = X_2 V S^{-1} W$ and their amplitudes $b$ obtained by fitting to the initial state of the dynamics, i.e., to the observables $x_i(0)$ at $t=0$. The final result is the spatio-temporal superposition $X \approx \Phi\,\Lambda^{m} b$, which reconstructs the data and can be also used to extrapolate the dynamics.
An example application to NEGF would be to treat the diagonal and subdiagonals of, say, $G^R(t,t')$ as different observables in constructing $X$, and extrapolate them with DMD in the diagonal time direction \cite{Yin2022}.

The above procedure integrates naturally into the QTT framework.
Without loss of generality, let us assume that $t$ is discretized according to Eq.~\eqref{eq:binary} such that we have $2^R$ time points.
To keep the number of columns equal to $2^R$ also in the time-shifted matrices, we append to them a zero column, i.e., $X_1=\bigl(\, x_i(0)\; x_i(dt)\; \cdots\; x_i(m\,dt)\; \cdots\; x_i(t_\mathrm{max}-dt)\; 0\,\bigr)$ and analogously for $X_2$.
This facilitates the representation of the time dependence stored in $X_1$ and $X_2$ in the quantics form
\begin{equation}
	\begin{aligned}
		X_1 &= \begin{tikzpicture}
	\begin{pgfonlayer}{nodelayer}
		\node [style=2-leg tensor] (3) at (0, 0.125) {};
		\node [style=none, label={below:$i$}] (10) at (-0.25, -0.4) {};
		\node [style=none, label={below:$m\vphantom{i}$}] (11) at (0.25, -0.4) {};
		\node [style=none] (12) at (-0.25, 0) {};
		\node [style=none] (13) at (0.25, 0) {};
	\end{pgfonlayer}
	\begin{pgfonlayer}{edgelayer}
		\draw [style=normal] (10.center) to (12.center);
		\draw [style=normal] (11.center) to (13.center);
	\end{pgfonlayer}
\end{tikzpicture} \!\approx \begin{tikzpicture}
	\begin{pgfonlayer}{nodelayer}
		\node [style=qtt] (1) at (-2, 0.125) {};
		\node [style=qtt] (2) at (-1, 0.125) {};
		\node [style=qtt] (3) at (0, 0.125) {};
		\node [style=none] (4) at (1, 0.125) {$\ldots$};
		\node [style=qtt] (5) at (2, 0.125) {};
		\node [style=none, label={below:$i$}] (8) at (-2, -0.4) {};
		\node [style=none, label={below:$t_1\vphantom{i}$}] (9) at (-1, -0.4) {};
		\node [style=none, label={below:$t_2\vphantom{i}$}] (10) at (0, -0.4) {};
		\node [style=none, label={below:$t_R\vphantom{i}$}] (12) at (2, -0.4) {};
		\node [style=none] (14) at (0.5, 0.125) {};
		\node [style=none] (15) at (1.5, 0.125) {};
	\end{pgfonlayer}
	\begin{pgfonlayer}{edgelayer}
		\draw [style=normal] (1) to (8.center);
		\draw [style=normal] (2) to (9.center);
		\draw [style=normal] (3) to (10.center);
		\draw [style=normal] (5) to (12.center);
		\draw [style=normal] (3) to (2);
		\draw [style=normal] (2) to (1);
		\draw [style=normal] (3) to (14.center);
		\draw [style=normal] (5) to (15.center);
	\end{pgfonlayer}
\end{tikzpicture}\\
		&\approx \begin{tikzpicture}
	\begin{pgfonlayer}{nodelayer}
		\node [style=qtt] (2) at (0.5, 0.125) {};
		\node [style=qtt] (3) at (1.5, 0.125) {};
		\node [style=none] (4) at (2.5, 0.125) {$\ldots$};
		\node [style=qtt] (5) at (3.5, 0.125) {};
		\node [style=none, label={below:$t_1\vphantom{i}$}] (9) at (0.5, -0.4) {};
		\node [style=none, label={below:$t_2\vphantom{i}$}] (10) at (1.5, -0.4) {};
		\node [style=none, label={below:$t_R\vphantom{i}$}] (12) at (3.5, -0.4) {};
		\node [style=none] (14) at (2, 0.125) {};
		\node [style=none] (15) at (3, 0.125) {};
		\node [style=qtt] (16) at (-3.5, 0.125) {};
		\node [style=qtt] (17) at (-2.5, 0.125) {};
		\node [style=none] (18) at (-1.5, 0.125) {$\ldots$};
		\node [style=qtt] (19) at (-0.5, 0.125) {};
		\node [style=none, label={below:$i_1\vphantom{i}$}] (20) at (-3.5, -0.4) {};
		\node [style=none, label={below:$i_2\vphantom{i}$}] (21) at (-2.5, -0.4) {};
		\node [style=none, label={below:$i_N\vphantom{i}$}] (22) at (-0.5, -0.4) {};
		\node [style=none] (23) at (-2, 0.125) {};
		\node [style=none] (24) at (-1, 0.125) {};
	\end{pgfonlayer}
	\begin{pgfonlayer}{edgelayer}
		\draw [style=normal] (2) to (9.center);
		\draw [style=normal] (3) to (10.center);
		\draw [style=normal] (5) to (12.center);
		\draw [style=normal] (3) to (2);
		\draw [style=normal] (3) to (14.center);
		\draw [style=normal] (5) to (15.center);
		\draw [style=normal] (16) to (20.center);
		\draw [style=normal] (17) to (21.center);
		\draw [style=normal] (19) to (22.center);
		\draw [style=normal] (17) to (16);
		\draw [style=normal] (17) to (23.center);
		\draw [style=normal] (19) to (24.center);
		\draw [style=normal] (19) to (2);
	\end{pgfonlayer}
\end{tikzpicture}\label{eq:X1qtt}
	\end{aligned}
\end{equation}
and, analogously, $X_2=\begin{tikzpicture}
	\begin{pgfonlayer}{nodelayer}
		\node [style=blackqtt] (2) at (2, 0.125) {};
		\node [style=none] (9) at (2, -0.4) {};
		\node [style=blackqtt] (16) at (-2, 0.125) {};
		\node [style=blackqtt] (17) at (-1, 0.125) {};
		\node [style=none] (18) at (0, 0.125) {$\ldots$};
		\node [style=blackqtt] (19) at (1, 0.125) {};
		\node [style=none] (20) at (-2, -0.4) {};
		\node [style=none] (21) at (-1, -0.4) {};
		\node [style=none] (22) at (1, -0.4) {};
		\node [style=none] (23) at (-0.5, 0.125) {};
		\node [style=none] (24) at (0.5, 0.125) {};
	\end{pgfonlayer}
	\begin{pgfonlayer}{edgelayer}
		\draw [style=normal] (2) to (9.center);
		\draw [style=normal] (16) to (20.center);
		\draw [style=normal] (17) to (21.center);
		\draw [style=normal] (19) to (22.center);
		\draw [style=normal] (17) to (16);
		\draw [style=normal] (17) to (23.center);
		\draw [style=normal] (19) to (24.center);
		\draw [style=normal] (19) to (2);
	\end{pgfonlayer}
\end{tikzpicture}$ with the same indices. In the first line above, we apply Eq.~\eqref{eq:binary} to factorize the index $m=t/dt=[t_1,\ldots,t_R]_2$, whereas in the second line we assume that the index $i$ also has some structure that allows decomposing it into subindices $(i_1,\ldots,i_N)$ and thus factorizing the corresponding tensor. This could be the binary encoding applied to $i$, but it can also be some other factorization, as the algorithm is general. 
 
When the incoming data set $X$ is decomposed into a QTT according to Eq.~\eqref{eq:X1qtt}, 
one can implement the zeroing of the last column and shifting by a time step, Eq.~\eqref{eq:X1X2matrices}, in the QTT representation.
The zeroing is done by subtracting from $X$ its projection on the time indices $t_n=1$, and the shifting by applying a low-rank tensor-train operator \cite{MSSTA}.
In the tensor diagrams below, we use the shorthand notation $i_n$, $t_n$ to label all corresponding legs, with $i_{\overline{n}}$, $t_{\overline{n}}$ meaning a reversed leg order. We also fix $N=R=3$ to simplify the diagrams (while the algorithm works for any $N$, $R$).

As mentioned, the DMD assumes that the time evolution $t\to t+dt$ is governed by a linear operator $A$,
\begin{equation}
	A X_1 = X_2,\quad A=X_2 X_1^{-1},
\end{equation}
whose eigendecomposition defines the dynamic modes and their complex frequencies. 
The advantage of tensor-based DMD is that the form \eqref{eq:X1qtt} is essentially constructed from a series of SVDs, making the calculation of $X_1^{-1}$ cheap \cite{Klus2018}. Assuming that we already have $X_1$ represented via \eqref{eq:X1qtt}, we transform it as
\begin{equation}
	X_1 = \begin{tikzpicture}
	\begin{pgfonlayer}{nodelayer}
		\node [style=leftorth] (0) at (-2.25, 0.125) {};
		\node [style=leftorth] (1) at (-1.25, 0.125) {};
		\node [style=2-leg tensor] (4) at (0, 0.125) {};
		\node [style=rightorth] (5) at (1.25, 0.125) {};
		\node [style=rightorth] (6) at (2.25, 0.125) {};
		\node [style=none] (7) at (-2.25, -0.4) {};
		\node [style=none] (8) at (-1.25, -0.4) {};
		\node [style=none] (12) at (1.25, -0.4) {};
		\node [style=none] (13) at (2.25, -0.4) {};
		\node [style=none] (14) at (-0.3, -0.4) {};
		\node [style=none] (15) at (0.3, -0.4) {};
		\node [style=none] (16) at (-0.3, 0.1) {};
		\node [style=none] (17) at (0.3, 0.1) {};
		\node [style=none, label={below:$\underbrace{\qquad\quad}_{i_n}$}] (18) at (-1.25, -0.25) {};
		\node [style=none, label={below:$\underbrace{\qquad\quad}_{t_n}$}] (19) at (1.25, -0.25) {};
	\end{pgfonlayer}
	\begin{pgfonlayer}{edgelayer}
		\draw [style=normal] (0) to (7.center);
		\draw [style=normal] (1) to (8.center);
		\draw [style=normal] (5) to (12.center);
		\draw [style=normal] (6) to (13.center);
		\draw [style=normal] (6) to (5);
		\draw [style=normal, in=360, out=180] (1) to (0);
		\draw [style=normal] (1) to (4);
		\draw [style=normal] (4) to (5);
		\draw [style=normal] (16.center) to (14.center);
		\draw [style=normal] (17.center) to (15.center);
	\end{pgfonlayer}
\end{tikzpicture} = \begin{tikzpicture}
	\begin{pgfonlayer}{nodelayer}
		\node [style=leftorth] (0) at (-1.875, 0.125) {};
		\node [style=leftorth] (1) at (-0.875, 0.125) {};
		\node [style=rightorth] (5) at (0.875, 0.125) {};
		\node [style=rightorth] (6) at (1.875, 0.125) {};
		\node [style=none] (7) at (-1.875, -0.4) {};
		\node [style=none] (8) at (-0.875, -0.4) {};
		\node [style=none] (12) at (0.875, -0.4) {};
		\node [style=none] (13) at (1.875, -0.4) {};
		\node [style=diamond] (19) at (0, 0.125) {};
		\node [style=leftorth] (22) at (-2.875, 0.125) {};
		\node [style=none] (24) at (-2.875, -0.4) {};
		\node [style=rightorth] (25) at (2.875, 0.125) {};
		\node [style=none] (26) at (2.875, -0.4) {};
		\node [style=none] (27) at (0, 1.025) {$S\vphantom{{}^\dagger}$};
		\node [style=none] (28) at (-1.875, 1.025) {$U\vphantom{{}^\dagger}$};
		\node [style=none] (29) at (1.875, 1.025) {$V^\dagger$};
		\node [style=none] (30) at (-0.4, 0.375) {${}_{\alpha}$};
		\node [style=none] (31) at (0.425, 0.375) {${}_{\beta}$};
		\node [style=none] (33) at (1.875, -0.825) {$\scriptstyle t_n$};
		\node [style=none] (34) at (-1.875, -0.8) {$\scriptstyle i_n$};
	\end{pgfonlayer}
	\begin{pgfonlayer}{edgelayer}
		\draw [style=normal] (0) to (7.center);
		\draw [style=normal] (1) to (8.center);
		\draw [style=normal] (5) to (12.center);
		\draw [style=normal] (6) to (13.center);
		\draw [style=normal] (6) to (5);
		\draw [style=normal, in=360, out=180] (1) to (0);
		\draw [style=normal] (22) to (24.center);
		\draw [style=normal] (25) to (26.center);
		\draw [style=normal] (22) to (0);
		\draw [style=normal] (1) to (19);
		\draw [style=normal] (19) to (5);
		\draw [style=normal] (6) to (25);
	\end{pgfonlayer}
\end{tikzpicture}\,,\label{eq:X1svd}
\end{equation}
where we first contract the two central tensors, then left-orthogonalize all tensors to the left of them, using QR or SVD, and right-orthogonalize all tensors to the right (denoted by the triangle shapes), and, finally, perform an SVD on the central tensor. Only the leading $r$ singular values are kept in the latter SVD, chosen by the tolerance $\tau_\mathrm{DMD}$, defined analogously as $\tau_\mathrm{cutoff}$ above. 
Importantly, the number $r$ will become the number of modes onto which we decompose the dynamics.
Denoting $U=\begin{tikzpicture}
	\begin{pgfonlayer}{nodelayer}
		\node [style=leftorth] (0) at (-0.5, 0.125) {};
		\node [style=leftorth] (1) at (0.5, 0.125) {};
		\node [style=none] (7) at (-0.5, -0.4) {};
		\node [style=none] (8) at (0.5, -0.4) {};
		\node [style=none] (19) at (1.125, 0.125) {};
		\node [style=leftorth] (22) at (-1.5, 0.125) {};
		\node [style=none] (24) at (-1.5, -0.4) {};
		\node [style=none] (25) at (1.125, 0.375) {${}_{\alpha}$};
		\node [style=none] (26) at (-0.1, -0.275) {$\scriptstyle i_n$};
	\end{pgfonlayer}
	\begin{pgfonlayer}{edgelayer}
		\draw [style=normal] (0) to (7.center);
		\draw [style=normal] (1) to (8.center);
		\draw [style=normal, in=360, out=180] (1) to (0);
		\draw [style=normal] (22) to (24.center);
		\draw [style=normal] (22) to (0);
		\draw [style=normal] (1) to (19.center);
	\end{pgfonlayer}
\end{tikzpicture}$, $S=\begin{tikzpicture}
	\begin{pgfonlayer}{nodelayer}
		\node [style=none] (1) at (-0.625, 0.125) {};
		\node [style=none] (5) at (0.625, 0.125) {};
		\node [style=diamond] (19) at (0, 0.125) {};
		\node [style=none] (20) at (-0.625, 0.35) {${}_{\alpha}$};
		\node [style=none] (22) at (0.65, 0.375) {${}_{\beta}$};
	\end{pgfonlayer}
	\begin{pgfonlayer}{edgelayer}
		\draw [style=normal] (1.center) to (19);
		\draw [style=normal] (19) to (5.center);
	\end{pgfonlayer}
\end{tikzpicture}$, $V^\dagger=\begin{tikzpicture}
	\begin{pgfonlayer}{nodelayer}
		\node [style=rightorth] (5) at (-0.525, 0.125) {};
		\node [style=rightorth] (6) at (0.475, 0.125) {};
		\node [style=none] (12) at (-0.525, -0.4) {};
		\node [style=none] (13) at (0.475, -0.4) {};
		\node [style=none] (19) at (-1.15, 0.125) {};
		\node [style=rightorth] (25) at (1.475, 0.125) {};
		\node [style=none] (26) at (1.475, -0.4) {};
		\node [style=none] (28) at (-1.15, 0.35) {${}_{\beta}$};
		\node [style=none] (29) at (0.975, -0.275) {$\scriptstyle t_n$};
	\end{pgfonlayer}
	\begin{pgfonlayer}{edgelayer}
		\draw [style=normal] (5) to (12.center);
		\draw [style=normal] (6) to (13.center);
		\draw [style=normal] (6) to (5);
		\draw [style=normal] (25) to (26.center);
		\draw [style=normal] (19.center) to (5);
		\draw [style=normal] (6) to (25);
	\end{pgfonlayer}
\end{tikzpicture}$\,, we have thus decomposed the whole tensor train $X_1$ as $X_1 = U S V^\dagger$, which is equivalent to the SVD of the $X_1$ matrix with an additional factorization of $U$, $V^\dagger$.
The pseudoinverse thus follows very simply as $X_1^{-1}=VS^{-1}U^\dagger=\begin{tikzpicture}
	\begin{pgfonlayer}{nodelayer}
		\node [style=rightorth] (0) at (-1.875, 0.125) {};
		\node [style=rightorth] (1) at (-0.875, 0.125) {};
		\node [style=leftorth] (5) at (0.875, 0.125) {};
		\node [style=leftorth] (6) at (1.875, 0.125) {};
		\node [style=none] (7) at (-1.875, -0.4) {};
		\node [style=none] (8) at (-0.875, -0.4) {};
		\node [style=none] (12) at (0.875, -0.4) {};
		\node [style=none] (13) at (1.875, -0.4) {};
		\node [style=diamondinv] (19) at (0, 0.125) {};
		\node [style=rightorth] (22) at (-2.875, 0.125) {};
		\node [style=none] (24) at (-2.875, -0.4) {};
		\node [style=leftorth] (25) at (2.875, 0.125) {};
		\node [style=none] (26) at (2.875, -0.4) {};
		\node [style=none] (27) at (0.4, 0.375) {${}_{\alpha}$};
		\node [style=none] (28) at (-0.375, 0.35) {${}_{\beta}$};
		\node [style=none] (30) at (2.325, -0.275) {$\scriptstyle i_{\overline{n}}$};
		\node [style=none] (31) at (-1.35, -0.3) {$\scriptstyle t_{\overline{n}}$};
	\end{pgfonlayer}
	\begin{pgfonlayer}{edgelayer}
		\draw [style=normal] (0) to (7.center);
		\draw [style=normal] (1) to (8.center);
		\draw [style=normal] (5) to (12.center);
		\draw [style=normal] (6) to (13.center);
		\draw [style=normal] (6) to (5);
		\draw [style=normal, in=360, out=180] (1) to (0);
		\draw [style=normal] (22) to (24.center);
		\draw [style=normal] (25) to (26.center);
		\draw [style=normal] (22) to (0);
		\draw [style=normal] (1) to (19);
		\draw [style=normal] (19) to (5);
		\draw [style=normal] (6) to (25);
	\end{pgfonlayer}
\end{tikzpicture}$ \cite{Klus2018}, where $S^{-1}=\begin{tikzpicture}
	\begin{pgfonlayer}{nodelayer}
		\node [style=none] (1) at (-0.625, 0.125) {};
		\node [style=none] (5) at (0.625, 0.125) {};
		\node [style=diamondinv] (19) at (0, 0.125) {};
		\node [style=none] (20) at (0.625, 0.35) {${}_{\alpha}$};
		\node [style=none] (21) at (-0.625, 0.325) {${}_{\beta}$};
	\end{pgfonlayer}
	\begin{pgfonlayer}{edgelayer}
		\draw [style=normal] (1.center) to (19);
		\draw [style=normal] (19) to (5.center);
	\end{pgfonlayer}
\end{tikzpicture}$ is easily computed since $S$ is diagonal.

While using the QTT form \eqref{eq:X1qtt} one could actually construct the high-dimensional operator $A$ directly, which is impossible with matrix-based DMD, it is still useful to focus on its low-rank approximation. This is in order to select only the most meaningful dynamic modes, avoiding fitting to noise, and also to facilitate a complete eigendecomposition. The rank-$r$ approximation to $A$, 
\begin{equation}
	\tilde{A} \equiv U^\dagger A U = U^\dagger X_2 V S^{-1},\label{eq:Atilde_def}
\end{equation}
can be computed in the tensor form as
\begin{gather}
	U^\dagger X_2 V = \begin{tikzpicture}
	\begin{pgfonlayer}{nodelayer}
		\node [style=blackqtt] (16) at (-2.5, 0.875) {};
		\node [style=blackqtt] (17) at (-1.5, 0.875) {};
		\node [style=blackqtt] (18) at (0.5, 0.875) {};
		\node [style=blackqtt] (19) at (2.5, 0.875) {};
		\node [style=blackqtt] (23) at (-0.5, 0.875) {};
		\node [style=blackqtt] (24) at (1.5, 0.875) {};
		\node [style=leftorth] (28) at (-1.5, -0.75) {};
		\node [style=leftorth] (29) at (-0.5, -0.75) {};
		\node [style=none, label={below:$\alpha$}] (31) at (-0.5, -1.275) {};
		\node [style=leftorth] (33) at (-2.5, -0.75) {};
		\node [style=rightorth] (35) at (0.5, -0.75) {};
		\node [style=rightorth] (36) at (1.5, -0.75) {};
		\node [style=none, label={below:$\beta$}] (37) at (0.5, -1.275) {};
		\node [style=rightorth] (40) at (2.5, -0.75) {};
		\node [style=none] (41) at (-1.075, 0) {${}_{i_n}$};
		\node [style=none] (42) at (1.15, 0) {${}_{t_n}$};
		\node [style=none] (43) at (-3.5, 0.875) {$X_2$};
		\node [style=none] (44) at (-3.5, -0.75) {$U^\dagger$};
		\node [style=none] (45) at (3.5, -0.75) {$V$};
	\end{pgfonlayer}
	\begin{pgfonlayer}{edgelayer}
		\draw [style=normal] (17) to (16);
		\draw [style=normal] (17) to (23);
		\draw [style=normal] (19) to (24);
		\draw [style=normal] (23) to (18);
		\draw [style=normal] (18) to (24);
		\draw [style=normal] (29) to (31.center);
		\draw [style=normal, in=360, out=180] (29) to (28);
		\draw [style=normal] (33) to (28);
		\draw [style=normal] (35) to (37.center);
		\draw [style=normal] (36) to (35);
		\draw [style=normal] (36) to (40);
		\draw [style=normal] (16) to (33);
		\draw [style=normal] (17) to (28);
		\draw [style=normal] (23) to (29);
		\draw [style=normal] (18) to (35);
		\draw [style=normal] (24) to (36);
		\draw [style=normal] (19) to (40);
	\end{pgfonlayer}
\end{tikzpicture} = \begin{tikzpicture}
	\begin{pgfonlayer}{nodelayer}
		\node [style=2-leg tensor] (3) at (0, 0.125) {};
		\node [style=none, label={below:$\alpha$}] (10) at (-0.25, -0.4) {};
		\node [style=none, label={below:$\beta$}] (11) at (0.25, -0.4) {};
		\node [style=none] (12) at (-0.25, 0) {};
		\node [style=none] (13) at (0.25, 0) {};
	\end{pgfonlayer}
	\begin{pgfonlayer}{edgelayer}
		\draw [style=normal] (10.center) to (12.center);
		\draw [style=normal] (11.center) to (13.center);
	\end{pgfonlayer}
\end{tikzpicture},\label{eq:UdagX2V}\\[0.5em]
	\tilde{A} = \begin{tikzpicture}
	\begin{pgfonlayer}{nodelayer}
		\node [style=2-leg tensor] (3) at (0, 0) {};
		\node [style=none, label={left:$\alpha\vphantom{'}$}] (10) at (-1, 0) {};
		\node [style=none] (12) at (-0.5, 0) {};
		\node [style=none] (13) at (0.5, 0) {};
		\node [style=none, label={right:$\alpha'$}] (15) at (2.75, 0) {};
		\node [style=diamondinv] (16) at (2, 0) {};
		\node [style=none] (17) at (0, 1) {$U^\dagger X_2 V$};
		\node [style=none] (18) at (2, 1) {$S^{-1}\vphantom{{}_2}$};
		\node [style=none] (19) at (1.225, 0.25) {${}_{\beta}$};
	\end{pgfonlayer}
	\begin{pgfonlayer}{edgelayer}
		\draw [style=normal] (10.center) to (12.center);
		\draw [style=normal] (16) to (15.center);
		\draw [style=normal] (13.center) to (16);
	\end{pgfonlayer}
\end{tikzpicture} = \begin{tikzpicture}
	\begin{pgfonlayer}{nodelayer}
		\node [style=2-leg tensor round] (3) at (0, 0.125) {};
		\node [style=none, label={[xshift=-1pt,yshift=1pt]below:$\alpha\vphantom{'}$}] (10) at (-0.25, -0.4) {};
		\node [style=none, label={[xshift=1pt,yshift=1pt]below:$\alpha'$}] (11) at (0.25, -0.4) {};
		\node [style=none] (12) at (-0.25, 0) {};
		\node [style=none] (13) at (0.25, 0) {};
	\end{pgfonlayer}
	\begin{pgfonlayer}{edgelayer}
		\draw [style=normal] (10.center) to (12.center);
		\draw [style=normal] (11.center) to (13.center);
	\end{pgfonlayer}
\end{tikzpicture}.\label{eq:Atilde_tensor}
\end{gather}
The indices $\alpha$, $\alpha'$, $\beta$ are of dimension $r$, which typically is $\lesssim 100$,  hence the final tensors in Eqs.~\eqref{eq:UdagX2V} and \eqref{eq:Atilde_tensor} are small $r \times r$ matrices. Performing contractions such as the one in Eq.~\eqref{eq:UdagX2V} in an appropriate sequence, which avoids intermediate tensors beyond order three, never creates any exponentially large objects while maintaining the exponentially fine resolution of the QTT format. This is an important advantage of QTT-DMD, since in DMD sufficient time resolution is required to correctly capture the dynamics of the known data and obtain reliable modes \cite{Pogorelyuk2018,Yin2022,Yin2023}.

Having obtained $\tilde{A}$, we eigendecompose it using standard dense-matrix routines,
\begin{equation}
	\tilde{A} = \begin{tikzpicture}
	\begin{pgfonlayer}{nodelayer}
		\node [style=2-leg tensor round] (3) at (0, 0.125) {};
		\node [style=none, label={[xshift=-1pt,yshift=1pt]below:$\alpha\vphantom{'}$}] (10) at (-0.25, -0.4) {};
		\node [style=none, label={[xshift=1pt,yshift=1pt]below:$\alpha'$}] (11) at (0.25, -0.4) {};
		\node [style=none] (12) at (-0.25, 0) {};
		\node [style=none] (13) at (0.25, 0) {};
	\end{pgfonlayer}
	\begin{pgfonlayer}{edgelayer}
		\draw [style=normal] (10.center) to (12.center);
		\draw [style=normal] (11.center) to (13.center);
	\end{pgfonlayer}
\end{tikzpicture} \!= \begin{tikzpicture}
	\begin{pgfonlayer}{nodelayer}
		\node [style=pentagon] (0) at (0, 0.125) {};
		\node [style=circle] (1) at (-1, 0.125) {};
		\node [style=circleinv] (2) at (1, 0.125) {};
		\node [style=none, label={[yshift=1pt]below:$\alpha\vphantom{'}$}] (5) at (-1, -0.4) {};
		\node [style=none, label={[yshift=1pt]below:$\alpha'$}] (7) at (1, -0.4) {};
		\node [style=none] (8) at (-0.45, 0.425) {${}_{\ell\vphantom{'}}$};
		\node [style=none] (9) at (0.55, 0.425) {${}_{\ell'}$};
		\node [style=none] (10) at (-1, 1.125) {$W\vphantom{{}^{-1}}$};
		\node [style=none] (11) at (0, 1.125) {$\Lambda\vphantom{{}^{-1}}$};
		\node [style=none] (12) at (1.4, 1.125) {$W^{-1}$};
	\end{pgfonlayer}
	\begin{pgfonlayer}{edgelayer}
		\draw [style=normal] (1) to (0);
		\draw [style=normal] (0) to (2);
		\draw [style=normal] (2) to (7.center);
		\draw [style=normal] (1) to (5.center);
	\end{pgfonlayer}
\end{tikzpicture}\!\!,
\end{equation}
where $\Lambda=\mathrm{diag}(\lambda_\ell)_{\ell=1}^r$ is a diagonal matrix of $r$ eigenvalues enumerated by $\ell$. 
The resulting eigenvectors, stored in the columns of $W$, are transformed into the dynamic modes, i.e., the eigenvectors of the unprojected operator $A$, by the contraction
\begin{equation}
	\begin{aligned}
		\Phi &= X_2 V S^{-1}W = \begin{tikzpicture}
	\begin{pgfonlayer}{nodelayer}
		\node [style=blackqtt] (16) at (-2.5, 0.875) {};
		\node [style=blackqtt] (17) at (-1.5, 0.875) {};
		\node [style=blackqtt] (18) at (0.5, 0.875) {};
		\node [style=blackqtt] (19) at (2.5, 0.875) {};
		\node [style=blackqtt] (23) at (-0.5, 0.875) {};
		\node [style=blackqtt] (24) at (1.5, 0.875) {};
		\node [style=none] (28) at (-1.5, 0.35) {};
		\node [style=none] (31) at (-0.5, 0.35) {};
		\node [style=none] (33) at (-2.5, 0.35) {};
		\node [style=rightorth] (35) at (0.5, -0.45) {};
		\node [style=rightorth] (36) at (1.5, -0.45) {};
		\node [style=rightorth] (40) at (2.5, -0.45) {};
		\node [style=none] (42) at (1.175, 0.175) {${}_{t_n}$};
		\node [style=none] (43) at (-3.5, 0.875) {$X_2$};
		\node [style=none] (45) at (3.325, -0.45) {$V$};
		\node [style=none, label={below:$\underbrace{\qquad\quad}_{i_n}$}] (46) at (-1.5, 0.7) {};
		\node [style=diamondinv] (49) at (0.5, -1.475) {};
		\node [style=none] (50) at (0.2, -1.95) {${}_{\alpha}$};
		\node [style=none] (51) at (0.225, -1) {${}_{\beta}$};
		\node [style=circle] (52) at (0.5, -2.45) {};
		\node [style=none, label={[yshift=1pt]below:$\ell$}] (53) at (0.5, -2.975) {};
		\node [style=none] (55) at (1.325, -2.425) {$W\vphantom{{}^{-1}}$};
		\node [style=none] (56) at (1.525, -1.45) {$S^{-1}$};
	\end{pgfonlayer}
	\begin{pgfonlayer}{edgelayer}
		\draw [style=normal] (17) to (16);
		\draw [style=normal] (17) to (23);
		\draw [style=normal] (19) to (24);
		\draw [style=normal] (23) to (18);
		\draw [style=normal] (18) to (24);
		\draw [style=normal] (36) to (35);
		\draw [style=normal] (36) to (40);
		\draw [style=normal] (16) to (33.center);
		\draw [style=normal] (17) to (28.center);
		\draw [style=normal] (18) to (35);
		\draw [style=normal] (24) to (36);
		\draw [style=normal] (19) to (40);
		\draw [style=normal] (23) to (31.center);
		\draw [style=normal] (52) to (53.center);
		\draw [style=normal] (35) to (49);
		\draw [style=normal] (52) to (49);
	\end{pgfonlayer}
\end{tikzpicture}\\[-1em]
		&= \begin{tikzpicture}
	\begin{pgfonlayer}{nodelayer}
		\node [style=blackqtt] (16) at (-2.5, 0.125) {};
		\node [style=blackqtt] (17) at (-1.5, 0.125) {};
		\node [style=blackqtt] (23) at (-0.5, 0.125) {};
		\node [style=none] (28) at (-1.5, -0.4) {};
		\node [style=none] (31) at (-0.5, -0.4) {};
		\node [style=none] (33) at (-2.5, -0.4) {};
		\node [style=none, label={below:$\underbrace{\qquad\quad}_{\displaystyle i_n}$}] (46) at (-1.5, -0.05) {};
		\node [style=circle] (52) at (0.5, 0.125) {};
		\node [style=none, label={[yshift=1pt]below:$\ell$}] (53) at (0.5, -0.4) {};
	\end{pgfonlayer}
	\begin{pgfonlayer}{edgelayer}
		\draw [style=normal] (17) to (16);
		\draw [style=normal] (17) to (23);
		\draw [style=normal] (16) to (33.center);
		\draw [style=normal] (17) to (28.center);
		\draw [style=normal] (23) to (31.center);
		\draw [style=normal] (52) to (53.center);
		\draw [style=normal] (23) to (52);
	\end{pgfonlayer}
\end{tikzpicture}.
	\end{aligned}
\end{equation}
If viewed as a matrix in $(i_n, \ell)$, the $\ell$-th column of $\Phi$ is simply an approximate eigenvector of the full operator $A$ corresponding to the approximate eigenvalue $\lambda_\ell$ \cite{Tu2014,Kaneko2025}. The ``spatial'' dependence of this mode is given by its dependence on $i_n$, whereas the temporal evolution is naturally encoded in the eigenvalue, which is a complex number $\lambda_\ell = |\lambda_\ell| e^{i\omega_\ell}$, $\omega_\ell \in \mathbb{R}$. 
Whenever a numerically obtained eigenvalue falls slightly outside the unit disk, we rescale it onto the unit circle to avoid unphysical exponentially growing contributions and recover the expected damped oscillatory dynamics.

The dynamic mode decomposition of the incoming data $X$ is simply a spatio-temporal superposition of the modes $\Phi$ with amplitudes $b$,
\begin{align}
	X \approx \Phi \Lambda^{m} b = \Phi \, \mathrm{diag}(|\lambda_\ell|^{m} e^{i\omega_\ell m}) \, b, \quad b = \Phi^{-1} x,\label{eq:dmd}
\end{align} 
where $m=t/dt = [t_1,\ldots,t_R]_2$ is the integer enumerating the time points, Eq.~\eqref{eq:binary}.
The above expression can be understood as repeatedly applying the operator $A$, representing the evolution $t\to t+dt$, to the first column of $X$, i.e., to the initial state of the dynamics, $x$. 
The latter is obtained by projecting $X$ onto the quantics indices $t_n=0$ and contracting the fixed tensors away, $x=\begin{tikzpicture}
	\begin{pgfonlayer}{nodelayer}
		\node [style=qtt] (25) at (-2, 0.125) {};
		\node [style=qtt] (26) at (-1, 0.125) {};
		\node [style=qtt] (29) at (0, 0.125) {};
		\node [style=none] (31) at (-1, -0.4) {};
		\node [style=none] (32) at (0, -0.4) {};
		\node [style=none] (33) at (-2, -0.4) {};
		\node [style=none] (34) at (-0.475, -0.275) {$\scriptstyle i_n$};
	\end{pgfonlayer}
	\begin{pgfonlayer}{edgelayer}
		\draw [style=normal] (26) to (25);
		\draw [style=normal] (26) to (29);
		\draw [style=normal] (25) to (33.center);
		\draw [style=normal] (26) to (31.center);
		\draw [style=normal] (29) to (32.center);
	\end{pgfonlayer}
\end{tikzpicture}$\,.
The amplitude vector $b$ is fitted to this initial condition $x$, with $\Phi^{-1}=\begin{tikzpicture}
	\begin{pgfonlayer}{nodelayer}
		\node [style=blackqtt] (16) at (-2.5, 0.15) {};
		\node [style=blackqtt] (17) at (-1.5, 0.15) {};
		\node [style=blackqtt] (23) at (-0.5, 0.15) {};
		\node [style=none] (28) at (-1.5, -0.375) {};
		\node [style=none] (31) at (-0.5, -0.375) {};
		\node [style=none] (33) at (-2.5, -0.375) {};
		\node [style=circle] (52) at (-3.5, 0.15) {};
		\node [style=none] (53) at (-3.5, -0.375) {};
		\node [style=none] (54) at (-3.1, -0.2) {$\scriptstyle \ell'$};
		\node [style=none] (55) at (-1, -0.275) {$\scriptstyle i_{\overline{n}}$};
	\end{pgfonlayer}
	\begin{pgfonlayer}{edgelayer}
		\draw [style=normal] (17) to (16);
		\draw [style=normal] (17) to (23);
		\draw [style=normal] (16) to (33.center);
		\draw [style=normal] (17) to (28.center);
		\draw [style=normal] (23) to (31.center);
		\draw [style=normal] (52) to (53.center);
		\draw [style=normal] (52) to (16);
	\end{pgfonlayer}
\end{tikzpicture}$ being the pseudoinverse of $\Phi$ treated as a matrix in ($i_n,\ell$). Explicitly, $b$ is thus obtained by the contraction
\begin{equation}
	b = \,\begin{tikzpicture}
	\begin{pgfonlayer}{nodelayer}
		\node [style=blackqtt] (16) at (-2.5, 0.525) {};
		\node [style=blackqtt] (17) at (-1.5, 0.525) {};
		\node [style=blackqtt] (23) at (-0.5, 0.525) {};
		\node [style=circle] (52) at (-3.525, 0.525) {};
		\node [style=none, label={below:$\scriptstyle \ell'$}] (53) at (-3.525, 0) {};
		\node [style=none] (55) at (-1.075, -0.125) {$\scriptstyle i_{\overline{n}}$};
		\node [style=qtt] (56) at (-2.5, -0.725) {};
		\node [style=qtt] (57) at (-1.5, -0.725) {};
		\node [style=qtt] (58) at (-0.5, -0.725) {};
		\node [style=none] (59) at (-4.55, 0.625) {$\Phi^{-1}$};
		\node [style=none] (60) at (0.25, -0.775) {$x$};
	\end{pgfonlayer}
	\begin{pgfonlayer}{edgelayer}
		\draw [style=normal] (17) to (16);
		\draw [style=normal] (17) to (23);
		\draw [style=normal] (52) to (53.center);
		\draw [style=normal] (52) to (16);
		\draw [style=normal] (57) to (56);
		\draw [style=normal] (57) to (58);
		\draw [style=normal] (56) to (16);
		\draw [style=normal] (57) to (17);
		\draw [style=normal] (58) to (23);
	\end{pgfonlayer}
\end{tikzpicture} = \begin{tikzpicture}
	\begin{pgfonlayer}{nodelayer}
		\node [style=hexagon] (52) at (-3.525, 0.15) {};
		\node [style=none, label={below:$\ell'$}] (53) at (-3.525, -0.375) {};
	\end{pgfonlayer}
	\begin{pgfonlayer}{edgelayer}
		\draw [style=normal] (52) to (53.center);
	\end{pgfonlayer}
\end{tikzpicture}
.
\end{equation}

The final step in evaluating Eq.~\eqref{eq:dmd} is to prepare $\Lambda^m$. This is not an obvious step because in order for Eq.~\eqref{eq:dmd} to produce $X$ as a complete QTT \eqref{eq:X1qtt} that describes the whole dynamics with exponentially fine resolution, we need a single tensor representation for $\Lambda^m$ containing all its $m$ powers. For a single eigenvalue $\lambda$, we note that
\begin{equation}
	\lambda^m = e^{m \ln{\lambda}}=e^{[t_1,\ldots,t_R]_2 \ln{\lambda}} = \prod_{n=1}^R e^{2^{R-n} t_n \ln{\lambda}} = \begin{tikzpicture}
	\begin{pgfonlayer}{nodelayer}
		\node [style=pentagon] (25) at (-1, 0.125) {};
		\node [style=pentagon] (26) at (0, 0.125) {};
		\node [style=pentagon] (29) at (1, 0.125) {};
		\node [style=none, label={below:$t_n$}] (31) at (0, -0.4) {};
		\node [style=none] (32) at (1, -0.4) {};
		\node [style=none] (33) at (-1, -0.4) {};
		\node [style=none] (35) at (0.5, 0.375) {\tiny $1$};
		\node [style=none] (36) at (-0.5, 0.375) {\tiny $1$};
	\end{pgfonlayer}
	\begin{pgfonlayer}{edgelayer}
		\draw [style=normal] (26) to (25);
		\draw [style=normal] (26) to (29);
		\draw [style=normal] (25) to (33.center);
		\draw [style=normal] (26) to (31.center);
		\draw [style=normal] (29) to (32.center);
	\end{pgfonlayer}
\end{tikzpicture},\label{eq:lambdalpowm}
\end{equation}	
which factorizes perfectly via Eq.~\eqref{eq:binary}, defining a simple $D=1$ QTT that can be directly prepared. To form the full $\Lambda^m$, we create one such QTT for each $\ell$ and label it by appending an additional tensor $\mathbb{I}_{\ell \Rightarrow l} = \begin{tikzpicture}
	\begin{pgfonlayer}{nodelayer}
		\node [style=onehot] (35) at (-2, 0.2) {};
		\node [style=none] (36) at (-2, -0.325) {};
		\node [style=none] (37) at (-1.75, -0.15) {$\scriptstyle\ell$};
	\end{pgfonlayer}
	\begin{pgfonlayer}{edgelayer}
		\draw [style=normal] (35) to (36.center);
	\end{pgfonlayer}
\end{tikzpicture}
$ whose only nonzero entry corresponds to setting the index $\ell$ to a specific value $l$. Then, we sum all these QTTs and obtain $\Lambda^m$. Diagrammatically,
\begin{equation}
	\begin{aligned}
	\Lambda^m &= \begin{tikzpicture}
	\begin{pgfonlayer}{nodelayer}
		\node [style=pentagon] (25) at (-1, 0.125) {};
		\node [style=pentagon] (26) at (0, 0.125) {};
		\node [style=pentagon] (29) at (1, 0.125) {};
		\node [style=none] (31) at (0, -0.4) {};
		\node [style=none] (32) at (1, -0.4) {};
		\node [style=none] (33) at (-1, -0.4) {};
		\node [style=none] (34) at (0.5, -0.275) {$\scriptstyle t_n$};
		\node [style=onehot] (35) at (-2, 0.125) {};
		\node [style=none, label={below:$\ell$}] (36) at (-2, -0.4) {};
		\node [style=none] (37) at (-2, 1) {$\mathbb{I}_{\ell \Rightarrow 1}$};
		\node [style=none] (38) at (0.5, 0.375) {\tiny $1$};
		\node [style=none] (39) at (-0.5, 0.375) {\tiny $1$};
		\node [style=none] (40) at (-1.575, 0.375) {\tiny $1$};
	\end{pgfonlayer}
	\begin{pgfonlayer}{edgelayer}
		\draw [style=normal] (26) to (25);
		\draw [style=normal] (26) to (29);
		\draw [style=normal] (25) to (33.center);
		\draw [style=normal] (26) to (31.center);
		\draw [style=normal] (29) to (32.center);
		\draw [style=normal] (35) to (36.center);
		\draw [style=normal] (35) to (25);
	\end{pgfonlayer}
\end{tikzpicture} ~~\,+ \begin{tikzpicture}
	\begin{pgfonlayer}{nodelayer}
		\node [style=pentagon] (25) at (-1, 0.125) {};
		\node [style=pentagon] (26) at (0, 0.125) {};
		\node [style=pentagon] (29) at (1, 0.125) {};
		\node [style=none] (31) at (0, -0.4) {};
		\node [style=none] (32) at (1, -0.4) {};
		\node [style=none] (33) at (-1, -0.4) {};
		\node [style=none] (34) at (0.5, -0.275) {$\scriptstyle t_n$};
		\node [style=onehot] (35) at (-2, 0.125) {};
		\node [style=none, label={below:$\ell$}] (36) at (-2, -0.4) {};
		\node [style=none] (37) at (-2, 1) {$\mathbb{I}_{\ell \Rightarrow 2}$};
		\node [style=none] (38) at (0.5, 0.375) {\tiny $1$};
		\node [style=none] (39) at (-0.5, 0.375) {\tiny $1$};
		\node [style=none] (40) at (-1.575, 0.375) {\tiny $1$};
	\end{pgfonlayer}
	\begin{pgfonlayer}{edgelayer}
		\draw [style=normal] (26) to (25);
		\draw [style=normal] (26) to (29);
		\draw [style=normal] (25) to (33.center);
		\draw [style=normal] (26) to (31.center);
		\draw [style=normal] (29) to (32.center);
		\draw [style=normal] (35) to (36.center);
		\draw [style=normal] (35) to (25);
	\end{pgfonlayer}
\end{tikzpicture} ~~\,+~~ \ldots\\[.5em]
	&= \begin{tikzpicture}
	\begin{pgfonlayer}{nodelayer}
		\node [style=pentagon] (25) at (-1, 0.125) {};
		\node [style=pentagon] (26) at (0, 0.125) {};
		\node [style=pentagon] (29) at (1, 0.125) {};
		\node [style=none, label={below:$t_n\vphantom{\ell}$}] (31) at (0, -0.4) {};
		\node [style=none] (32) at (1, -0.4) {};
		\node [style=none] (33) at (-1, -0.4) {};
		\node [style=blackcircle] (35) at (-2, 0.125) {};
		\node [style=none, label={below:$\ell$}] (36) at (-2, -0.4) {};
	\end{pgfonlayer}
	\begin{pgfonlayer}{edgelayer}
		\draw [style=normal] (26) to (25);
		\draw [style=normal] (26) to (29);
		\draw [style=normal] (25) to (33.center);
		\draw [style=normal] (26) to (31.center);
		\draw [style=normal] (29) to (32.center);
		\draw [style=normal] (35) to (36.center);
		\draw [style=normal] (35) to (25);
	\end{pgfonlayer}
\end{tikzpicture} \,~\to \begin{tikzpicture}
	\begin{pgfonlayer}{nodelayer}
		\node [style=pentagon] (25) at (-1, 0.125) {};
		\node [style=pentagon] (26) at (0, 0.125) {};
		\node [style=pentagon] (29) at (1, 0.125) {};
		\node [style=none, label={below:$t_n\vphantom{\ell}$}] (31) at (0, -0.4) {};
		\node [style=none] (32) at (1, -0.4) {};
		\node [style=none] (33) at (-1, -0.4) {};
		\node [style=blackcircle] (35) at (-2, 0.125) {};
		\node [style=none, label={below:$\ell$}] (36) at (-2, -0.4) {};
		\node [style=none, label={left:$\ell'\!$}] (37) at (-2.7, 0.125) {};
	\end{pgfonlayer}
	\begin{pgfonlayer}{edgelayer}
		\draw [style=normal] (26) to (25);
		\draw [style=normal] (26) to (29);
		\draw [style=normal] (25) to (33.center);
		\draw [style=normal] (26) to (31.center);
		\draw [style=normal] (29) to (32.center);
		\draw [style=normal] (35) to (36.center);
		\draw [style=normal] (35) to (25);
		\draw [style=normal] (37.center) to (35);
	\end{pgfonlayer}
\end{tikzpicture}\,,
	\end{aligned}\label{eq:Lambdapowm}
\end{equation}
where in the last step we added an index $\ell'$ (by contracting with a $\delta_{\ell\ell'\ell''}$ tensor) such that the tensor train represents powers of a \emph{diagonal $\Lambda^m$ matrix}. We perform the above sum exactly with the direct-sum algorithm.

The algorithm concludes by performing the contraction \eqref{eq:dmd}, 
\begin{equation}
	\begin{aligned}
		X &\approx\, \begin{tikzpicture}
	\begin{pgfonlayer}{nodelayer}
		\node [style=blackqtt] (16) at (-2.5, 0.125) {};
		\node [style=blackqtt] (17) at (-1.5, 0.125) {};
		\node [style=blackqtt] (23) at (-0.5, 0.125) {};
		\node [style=none, label={below:$i_n$}] (28) at (-1.5, -0.4) {};
		\node [style=none] (31) at (-0.5, -0.4) {};
		\node [style=none] (33) at (-2.5, -0.4) {};
		\node [style=circle] (52) at (0.5, 0.125) {};
		\node [style=pentagon] (54) at (2.5, 0.125) {};
		\node [style=pentagon] (55) at (3.5, 0.125) {};
		\node [style=pentagon] (56) at (4.5, 0.125) {};
		\node [style=none, label={below:$t_n\vphantom{i}$}] (57) at (3.5, -0.4) {};
		\node [style=none] (58) at (4.5, -0.4) {};
		\node [style=none] (59) at (2.5, -0.4) {};
		\node [style=blackcircle] (60) at (1.5, 0.125) {};
		\node [style=hexagon] (63) at (1.5, -1) {};
		\node [style=none] (65) at (1.025, 0.45) {$\scriptstyle\ell$};
		\node [style=none] (66) at (1.875, -0.4) {$\scriptstyle\ell'$};
		\node [style=none] (67) at (-1, 1) {$\Phi$};
		\node [style=none] (68) at (1.5, -1.75) {$b$};
		\node [style=none] (69) at (3.25, 1) {$\Lambda^m$};
	\end{pgfonlayer}
	\begin{pgfonlayer}{edgelayer}
		\draw [style=normal] (17) to (16);
		\draw [style=normal] (17) to (23);
		\draw [style=normal] (16) to (33.center);
		\draw [style=normal] (17) to (28.center);
		\draw [style=normal] (23) to (31.center);
		\draw [style=normal] (23) to (52);
		\draw [style=normal] (55) to (54);
		\draw [style=normal] (55) to (56);
		\draw [style=normal] (54) to (59.center);
		\draw [style=normal] (55) to (57.center);
		\draw [style=normal] (56) to (58.center);
		\draw [style=normal] (60) to (54);
		\draw [style=normal] (52) to (60);
		\draw [style=normal] (60) to (63);
	\end{pgfonlayer}
\end{tikzpicture}.\label{eq:dmdqtt}\\[.5em]
		&= \begin{tikzpicture}
	\begin{pgfonlayer}{nodelayer}
		\node [style=blackqtt] (16) at (-2.5, 0.125) {};
		\node [style=blackqtt] (17) at (-1.5, 0.125) {};
		\node [style=blackqtt] (23) at (-0.5, 0.125) {};
		\node [style=none, label={below:$i_n$}] (28) at (-1.5, -0.4) {};
		\node [style=none] (31) at (-0.5, -0.4) {};
		\node [style=none] (33) at (-2.5, -0.4) {};
		\node [style=pentagon] (54) at (0.5, 0.125) {};
		\node [style=pentagon] (55) at (1.5, 0.125) {};
		\node [style=pentagon] (56) at (2.5, 0.125) {};
		\node [style=none, label={below:$t_n\vphantom{i}$}] (57) at (1.5, -0.4) {};
		\node [style=none] (58) at (2.5, -0.4) {};
		\node [style=none] (59) at (0.5, -0.4) {};
	\end{pgfonlayer}
	\begin{pgfonlayer}{edgelayer}
		\draw [style=normal] (17) to (16);
		\draw [style=normal] (17) to (23);
		\draw [style=normal] (16) to (33.center);
		\draw [style=normal] (17) to (28.center);
		\draw [style=normal] (23) to (31.center);
		\draw [style=normal] (55) to (54);
		\draw [style=normal] (55) to (56);
		\draw [style=normal] (54) to (59.center);
		\draw [style=normal] (55) to (57.center);
		\draw [style=normal] (56) to (58.center);
		\draw [style=normal] (23) to (54);
	\end{pgfonlayer}
\end{tikzpicture}\,.
	\end{aligned}
\end{equation}
The above can be understood as a certain ``recompression'' of the original QTT data $X$, where the dimension of the central bond is now given by the number $r$ of the dynamic modes used (prior to the usual truncation). 
Since the sum of exponentials \eqref{eq:Lambdapowm} is smooth, this recompression can also denoise the dynamics via a suitably chosen tolerance $\tau_\mathrm{DMD}$ within the SVD \eqref{eq:X1svd}.
We also note that while Eq.~\eqref{eq:negfqtt} applies interleaved ordering to the indices $t_n$, $t'_n$, Eqs.~\eqref{eq:X1qtt} and \eqref{eq:dmdqtt} use a sequential ordering of $i_n$, $t_n$. Since the sequential ordering is necessary for the QTT-DMD algorithm, usually the first step is to preprocess the data from the interleaved to the sequential form, e.g., by a means of several pair-wise swaps using SVD. A low-rank sequential ordering is typically the coarse-centered order, $(i_N,\ldots,i_1,t_1,\ldots,t_R)$, which we also use in our applications.

Equation \eqref{eq:dmdqtt} decomposes the original data $X$ onto the same time grid that it started with, $t\in[0,t_\mathrm{max}]$. However, the time $t$ in the dynamic mode decomposition is a parameter that can be continuously varied and one can easily cast the data onto a new grid. In particular, an extrapolation or interpolation can be performed. To this end, when preparing $\Lambda^m$ according to Eqs.~\eqref{eq:lambdalpowm}, \eqref{eq:Lambdapowm}, one simply increases $R$ by one, adding one more $t_n$ tensor. If the added tensor is understood as a coarse scale, i.e., it is added to the left of existing $t_n$, then one extrapolates the data onto a doubled time domain. In the opposite case of adding a fine scale tensor, one interpolates the data, increasing the resolution by a factor of two.

\subsection{Predictor-corrector solver with block-time-stepping update}\label{predictor-corrector}

\subsubsection{General procedure}

When solving the KBE with functions represented in a compressed form, here as QTTs, it is most straightforward to solve them globally. That is, at each self-consistent iteration, one updates all $(t,t')$ time points simultaneously.
Within Eqs.~\eqref{eq:ret}-\eqref{eq:les} this corresponds to the matrix inversion of the $(1-G_0^R\cdot\Sigma^R\cdot{})$ operator. While convenient implementation-wise, this approach suffers from instabilities or trapping of the self-consistent iterations which update the self-energy, if the starting point is too far from the solution \cite{Sroda2024,Gasperlin2025}. 
For instance, for large $t_\mathrm{max}$ and strong enough interaction (e.g., $t_\mathrm{max} \gtrsim 100$, interaction $U \gtrsim 1$ for the Hubbard model with bandwidth 8), this occurs when starting the iteration from the noninteracting Green's function $G_{0\mathbf{k}}$.
While the issue can be cured to some extent by slowly ramping up the interaction $U$ 
during the iteration \cite{Sroda2024}, for large enough $U$ and $t_\mathrm{max}$, the ramping would need to be too slow for efficient calculations.

The above convergence problems are not a result of the compressed format, but simply due to the properties of self-consistent iterations of the non-Markovian KBE.
The features at long times, $t, t' \to t_\mathrm{max}$, are determined by an integral over the whole system's past and hence they cannot converge well until the short times do.
As such, even in late iterations, small improvements of the solution around $t,t' \to 0$ may affect all future times, leading to convergence plateaus \cite{Inayoshi2025}.
These problems are bypassed when taking advantage of causality to solve the equations in a time-stepping fashion \cite{Nessi,stefanucci_van_leeuwen_2013}. For instance, the NESSi package \cite{Nessi} solves the KBE one time step at a time, with the self-consistency at a given time step started from a polynomial extrapolation of past data (predictor step) and then improved in a small number of iterations (corrector steps). The equations on this single time step are stable and converge quickly, because one keeps the past time step solutions fixed and starts from a good prediction. 

Let us adapt the above predictor-corrector method to the QTT framework.
The underlying observation is that the global iteration can be stable provided that the domain being solved is small enough.
Crucially, it is the physical time extent of the domain that matters and not how fine the contour is discretized.
For instance, a global iteration with any fixed $t_\mathrm{max}$ is as difficult to converge with $dt = 10^{-6}$ as with $dt = 10^{-8}$, even though the latter involves $10^2$ times more time points. 
In the QTT form \eqref{eq:negfqtt}, also the bond dimension $D$, which defines the storage requirement and computational complexity, is indifferent to varying $dt$ in this range.
It is thus possible to solve the KBE globally on consecutive subdomains of physical time extent short enough to stabilize the iteration, but orders of magnitude larger than in conventional matrix-based NEGF calculations. Also the very fine quantics resolution can be fully retained.

\begin{figure*}[p]
	\centering
	\includegraphics[height=.91\textheight]{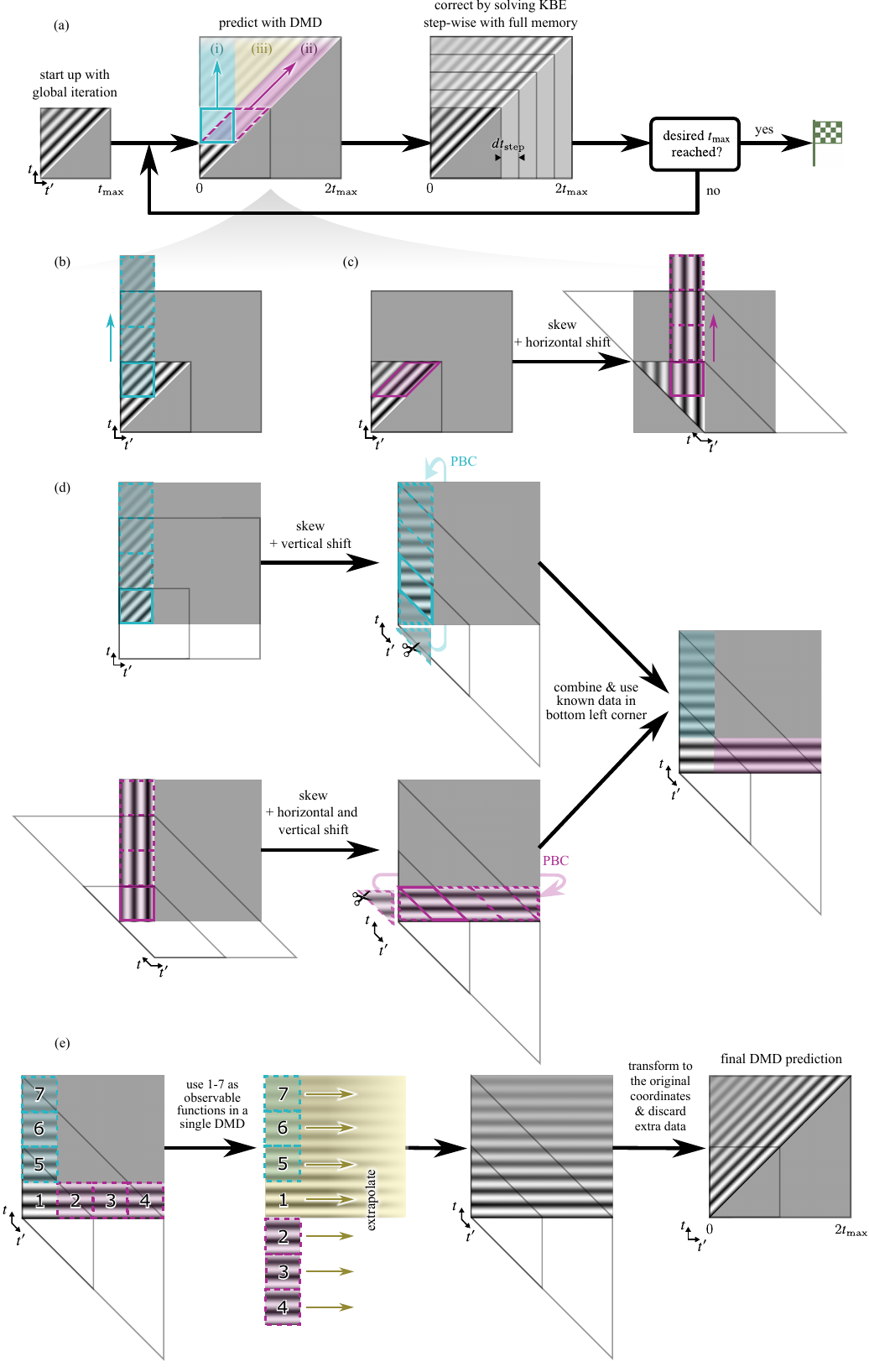}
	\caption{(a) Block time stepping solution of the Kadanoff-Baym equations. (b)-(e) Details of the application of the QTT-DMD algorithm to extrapolate NEGFs. 
	In panels (b)-(e), the dark grey square represents the full square window represented by the stored QTTs, whereas the faint grey lines correspond to the way the original time coordinates are modified by the affine transformations.
	We use periodic boundary conditions in the latter transformations, hence when a data set extends beyond the window, it is wrapped back into it.}
	\label{fig:sketch}
\end{figure*}

The procedure is termed \emph{block time stepping} \cite{Inayoshi2025} and we summarize it in Fig.~\ref{fig:sketch}(a), with the details represented in panels (b)-(e). It proceeds as follows.
After obtaining the Matsubara-axis solution, we start the iteration on the real-time axis with a modest and stable $t_\mathrm{max}\sim 10$, for which we solve the KBE globally with $G_{0\mathbf{k}}$ as the ansatz.
Next, we double $t_\mathrm{max}$ and fill the extended part of the domain with a prediction obtained from DMD. 
This prediction is then corrected one by one in large \emph{block time steps}, $dt_\mathrm{step} \sim 4$, which for a typical $dt_\mathrm{step} \sim 0.01$ in a matrix-based NEGF calculation would correspond to 400 conventional time steps.
Note that our time steps in practice contain a huge number of grid points $dt_\mathrm{step}/dt \sim 10^6$, due to the very fine resolution of our time grid.
Once the doubled domain is fully converged, the calculation either ends, if we reached our final $t_\mathrm{max}$, or another set of doubling, prediction and correction is repeated.
In the following, we provide more details on the prediction and correction steps.

\subsubsection{Prediction}

For concreteness, consider extrapolating $G_\mathbf{k}^R$ at a fixed $\mathbf{k}$ point.
As preprocessing, we double the domain from $[0,t_\mathrm{max}]^2$ to $[0,2t_\mathrm{max}]^2$ by appending two coarse tensors, $t_n$, $t'_n$ to the left of the input QTT, and filling the enlarged domain with zeros.
Then, we proceed in three steps in order to use as input to DMD the data closest to the converged $t_\mathrm{max}$.

The first is to extrapolate the decay along $t$, as shown in Fig.~\ref{fig:sketch}(b).
To this end, we take the top left square of the known data, corresponding to fixing the first four tensors $[t_1,t'_1,t_2,t'_2,\ldots]_2$ of the enlarged QTT to $[0,0,1,0,\ldots]_2$.
After contracting the fixed tensors away, the data is presented to QTT-DMD with $t$ understood as the direction of time evolution and $t'$ as different observable functions.
The output of the QTT-DMD is a QTT with two additional coarse $t_n$ tensors, corresponding to extrapolating into a domain four times that of the input data.

The second step is to perform an extrapolation along the diagonal direction, see Fig.~\ref{fig:sketch}(c).
We apply an affine transformation to skew the original data into a parallelogram (faint gray lines) such that the diagonals are oriented vertically, and we also apply a translation so that the parallelogram fits within the enlarged domain (solid gray square).
Note that the output of the affine transformation is also a square matrix, with the data that falls out of the square ``window'' being wrapped around via periodic boundary conditions.
In the sketches, the gray lines always represent how the original coordinates are transformed, whereas the dark gray box is the square ``window'' one obtains after transformations.
The affine transformation is performed as a contraction with a low-$D$ tensor-train operator \cite{MSSTA}, constructed according to the algorithm in Appendix B of Ref.~\cite{Rohshap2025}.
As input data to QTT-DMD, we take the square subdomain obtained by fixing the tensors to $[0,0,1,1,\ldots]_2$, which in the untransformed space corresponds to a parallelogram aligned along the diagonals.
Then, we extrapolate the same way as before.

At this point, we have two QTTs: one describing the vertical extrapolation in the untransformed space (marked by four blue squares in Fig.~\ref{fig:sketch}(b)) and one describing the diagonal extrapolation in the skewed space (marked by four violet squares in Fig.~\ref{fig:sketch}(b)).

The final step is to fill in the data in the wedge between the two extrapolations (yellow shaded region in Fig.~\ref{fig:sketch}(a)).
We start by preprocessing the above extrapolations.
We embed both of the QTTs such that they represent a column within an enlarged domain (dark gray square in Fig.~\ref{fig:sketch}(d)).
Next, we affine transform both into a vertical skew, which allows us to represent the entire triangular domain of the $G_\mathbf{k}^R$ within one square window, with the diagonals lying horizontally and the off-diagonals vertically.
We then combine (sum) these transformed QTTs and also a corner of the known converged data and obtain an L-shaped domain (far right of Fig.~\ref{fig:sketch}(d)), which concludes the preprocessing.

Finally, in order to use all the information from within the L-shaped domain, and extrapolate the data such that discontinuities at the domain's edges are minimized, we apply a technique akin to higher-order DMD \cite{LeClainche2017,Pan2020,Schmid2022,Yin2022,Yin2023,Kaneko2025}. That is, we split the L domain into 7 patches (Fig.~\ref{fig:sketch}(e)), which we treat as observable functions. These functions are then extrapolated along the horizontal direction (diagonal in the original coordinates). For patches 1 and 5-7 this is equivalent to the usual DMD. However, the addition of patches 2-4 serves as a time-delay embedding \cite{LeClainche2017,Pan2020}, which provides additional information about the dynamical system. As a result, the fitting produces a continuous extrapolation within the whole square window.
This is actually more data than necessary, as only the lower triangle exists in our original coordinates.
A final affine transform brings us back to the original coordinates, concluding the extrapolation procedure.
Note that at the end we place the known converged data into the $[0,t_\mathrm{max}]^2$ subdomain.

Importantly, to give the DMD even more information about the dynamical system, we extrapolate all $R$, $<$, $\rceil$ components simultaneously within the above procedure.
To this end, we add an observable-function tensor to the DMD input, which distinguishes between the components (for $G_\mathbf{k}^\rceil$ in Fig.~\ref{fig:sketch}(b) we use its $[t_\mathrm{max}/2,t_\mathrm{max}] \times [0,\beta]$ subdomain). 
Concretely, in Fig.~\ref{fig:sketch}(b), we simultaneously extrapolate all three components, while in (c) and (d) extrapolate the two real-time ones.
This extrapolation procedure helps to preserve the proper boundary condition of the Green's function \cite{Aoki2014}.
A small technicality arises when the number of time points on the imaginary and real-time axes is different.
Then, to combine the components into a single QTT, we need to upsample either the imaginary or the real-time axis, such that there is the same number of observable functions in both components.
This does not affect the extrapolation, as the upsampling is aiming to artificially increase the number of observable functions, and not the resolution of the time axis along which the DMD extrapolates.
At the end, we discard the upsampled observables.
As for the lesser component, we make use of a symmetry relation \cite{Aoki2014} to obtain the second triangular domain.

One might wonder why we do not simply take the whole $[0,t_\mathrm{max}]^2$ domain, treat the direction $t$ as the observables, and extrapolate along $t'$ using the whole information from the known square. The first reason is that this obviously does not work for the retarded component (if explicitly defined with $\theta_\mathcal{C}(t,t')$ as here).
Furthermore, for the lesser component, we find that such an extrapolation leads to the revival of the diagonal peak, which is clearly an artifact, as the Green's function is expected to decay away from the diagonal.

Another possible tweak to our scheme would be to use the cross extrapolation \cite{Jeannin2024} to extrapolate data from the L-shaped domain in Fig.~\ref{fig:sketch}(e).
We explored this idea by incorporating the cross extrapolation into the tensor-cross-interpolation sweeping algorithm \cite{Fernandez2024} and found it very sensitive to small variations in the data, especially if the variations have different origins in different parts of the domain.
In our L-shaped domain, the vertical data comes from one DMD extrapolation, the horizontal from another, and the corner is the well-converged input data.
Using the subdomains containing DMD-extrapolated data as input data to the cross extrapolation algorithm resulted in numerical problems and we found it to be impracticable.
Similar issues were reported in Ref.~\cite{Jeannin2024}, where using the cross-extrapolated data as input to a subsequent cross extrapolation also led to numerical instabilities.

\subsubsection{Correction}

The easiest approach would be to correct the prediction in one big step involving only the doubled part. However, depending on the physical problem, doubling involves a solution on a big enough domain that one still suffers from slow convergence, as in a global iteration.

In order to avoid sacrificing our very fine grid spacing, $dt < 10^{-6}$, we do not update individual times, but rather update the solution on \emph{block time steps} \cite{Inayoshi2025} with width $d t_\mathrm{step} \gg dt$ encompassing many time steps of width $dt$. 
Here, the block time steps have an L shape (Fig.~\ref{fig:sketch}(a)) and are represented by a single QTT in the full $[0,2t_\mathrm{max}]^2$ domain with $(t,t')$ external to the L zeroed out, corresponding to a variant of the procedure of Ref.~\cite{Inayoshi2025}.
The zeroes cost very little in the QTT representation, especially when the steps are chosen such that they conform to the underlying bisection-like quantics grid, which minimizes the effect of the discontinuities on the bond dimensions.
This requires that the number of steps is a power of two, which we assume to be the case here.
We also allow each doubling to involve a different number of block time steps, as for different stages of the dynamics a stable solution of the KBE might require smaller or larger $dt_\mathrm{step}$.

Another advantage of block time stepping is that to calculate the diagrams we can use the QTTs restricted to the blocks,
as the diagrams are given by element-wise products. This is very beneficial, as it strongly reduces the bond dimensions, since the QTT representing a block time step contains a much simpler time dependence than the whole correlator.
Hence, even though depending on the value of $dt_\mathrm{step}$, we might need a rather large number of self-consistent iterations, each iteration is much cheaper than a global one, resulting in a performance gain.
A similar approach could be used for convolutions \cite{Inayoshi2025}, by treating them as a multiplication of block matrices. However, for simplicity, we calculate the convolutions by first merging the past data with the given step and performing a single matrix multiplication. Similarly, the Dyson equation is solved by merging the past (well converged) data with the time step being solved (while the future steps are zeroed out until the current one is fully converged). 
Since we use a DMRG-like solver \cite{Sroda2024}, which works by optimizing an initial QTT ansatz, which in the present case is well converged everywhere except on the rather thin block time step (with a good initial guess of the solution), we are satisfied with the performance of this scheme.
Future implementations, however, should calculate convolutions and solve the Dyson equation only on the block time step for increased performance \cite{Inayoshi2025}.

To summarize, the correction procedure is basically the same solution of the KBE as we introduced in Ref.~\cite{Sroda2024}, but with the diagram calculations restricted to the step and the past (well converged) self-energy being kept fixed. This not only speeds up the calculation, but enforces causality and stabilizes the self-consistency.
We stress that we always use the full memory kernel in these calculations, unlike in memory truncation schemes \cite{Schueler2018,Stahl2022}.

\section{Results}\label{results}

In this section, we show results for an interaction quench in the Hubbard model on a $32 \times 32$ square lattice. The Hamiltonian reads
\begin{equation}
	H(t) = \sum_{\mathbf{k}\sigma} \epsilon_{\mathbf{k}} c^\dagger_{\mathbf{k}\sigma} c_{\mathbf{k}\sigma} + U(t) \sum_i (n_{i\uparrow}-\tfrac12)(n_{i\downarrow}-\tfrac12).
\end{equation}
Here, $c^\dagger_{\mathbf{k}\sigma}$ creates an electron with momentum $\mathbf{k}$ and spin $\sigma$, $\epsilon_{\mathbf{k}}=-2t_\mathrm{h}(\cos{k_x}+\cos{k_y})$ is the electron dispersion with the hopping amplitude $t_\mathrm{h}$, and $n_{i\sigma}$ is the particle-number operator for spin $\sigma$ and site $i$. We set $t_\mathrm{h}=1$, express time in units of $1/t_\mathrm{h}$ ($\hbar\equiv 1$), and consider paramagnetic states, suppressing the spin index hereafter. The electronic correlations are treated within the self-consistent $GW$ approximation \cite{Sroda2024}.

\begin{figure}[t]
	\centering
	\includegraphics[]{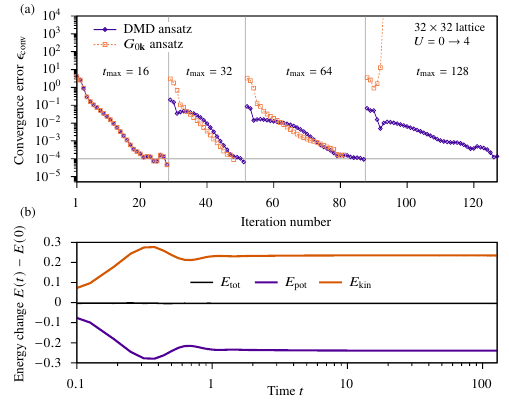}
	\caption{Comparison of the stability of a self-consistent iteration using the DMD prediction or the noninteracting Green's function $G_{0\mathbf{k}}$ as an initial guess. The simulation is for a $U=0$ to $U=4$ quench in a $32 \times 32$ Hubbard lattice at inverse temperature $\beta=100$. (a) Convergence error $\epsilon_\mathrm{conv}$ vs the iteration number, with each point being a single self-consistency iteration. The gray horizontal line marks the convergence goal, $10^{-4}$, whereas the gray vertical lines mark the iterations where the contour is doubled, with the initial contour length being $t_\mathrm{max}=16$ and the final one being $t_\mathrm{max}=128$. Note that each successive doubling requires more iterations to converge, which is due to the growing size of the domain that is being updated. (b) Time evolution of the variation in the potential energy  $E_\mathrm{pot}(t)-E_\mathrm{pot}(0)$, kinetic energy $E_\mathrm{kin}(t)-E_\mathrm{kin}(0)$, and total energy $E_\mathrm{tot}(t)-E_\mathrm{tot}(0)$ for the final converged solution from (a) that used the DMD ansatz.}
	\label{fig:stability}
\end{figure}

One of our main results is the stabilization of the QTT solver of the KBE.
In Fig.~\ref{fig:stability}, we investigate the convergence properties for a simulation of a sudden quench from $U=0$ to $U=4$, using both the noninteracting $G_{0\mathbf{k}}$ and the DMD extrapolation as initial guess.
The calculation involved 4 doublings, starting from a global start-up procedure on a time interval of length $t_\mathrm{max}=16$.
Within each doubling, there is only one block time step, i.e., the whole extended domain is solved for at once (with the past domain fixed).
The convergence goal is set to $10^{-4}$ \cite{Sroda2024}, with the convergence error between subsequent iterations defined as 
$
\epsilon_\mathrm{conv} = \mathrm{max}_\mathbf{k} \sum_{\alpha=R,\rceil,\gtrless} {|G^\alpha_{\mathbf{k},\mathrm{new}}-G_{\mathbf{k},\mathrm{old}}^{\alpha}|_\mathrm{F}}/{|G_{\mathbf{k},\mathrm{old}}^\alpha|_\mathrm{F}}\label{eq:convprec}
$.
In this estimate, we assume that $|G_{\mathbf{k},\mathrm{new}}^\alpha|_\mathrm{F} \approx |G_{\mathbf{k},\mathrm{old}}^\alpha|_\mathrm{F}$, which should be the case close to convergence, and we calculate the norms on the whole domain (including both the converged past data and the current block time step).
To more symmetrically measure convergence of the occupied and unoccupied $\mathbf{k}$ points, we average the errors for the lesser and greater components into one value with weights given by the components' respective norms.
This and subsequent simulations use a truncation cutoff of $\tau_\mathrm{cutoff}=10^{-11}$ in the tensor-network operations, except the SVD cutoff defining the number of DMD modes (Eq.~\eqref{eq:X1svd}) which is set to $\tau_\mathrm{DMD}=10^{-4}$. 
Crucially, we choose the latter cutoff to be larger than the square of the convergence error, i.e., $\tau_\mathrm{DMD} > \epsilon_\mathrm{conv}^2$, where the square follows from the different ways we define the convergence error and the SVD tolerance (see above).
Otherwise, the DMD tries to learn dynamics from noise.
In the two-site DMRG-like procedure that solves the linear problem Eqs.~\eqref{eq:mat}-\eqref{eq:les}, we use 2 sweeps with 2-4 local Krylov iterations.
The initial Matsubara-axis and the start-up global solutions are converged to $5 \times 10^{-5}$.
To speed up the calculations, when converging a block time step, we often increase the maximum bond dimension from $D \sim 20{-}50$ to unrestricted or to $150$.

Clearly, the DMD prediction performs better than simply starting from $G_{0\mathbf{k}}$ [Fig.~\ref{fig:stability}(a)].
While in the first three doublings, both ans\"atze perform similarly, the DMD remains stable in the last one, where using $G_{0\mathbf{k}}$ leads to a divergence.
While it might be possible to fine tune the calculation with the $G_{0\mathbf{k}}$ ansatz and get it to converge (e.~g. using interaction ramps or linear mixing), the calculation with the DMD ansatz is inherently more stable.
Note that running this simulation with a fully global iteration (initialized with $G_{0\mathbf{k}}$) leads to an instability already for $t_\mathrm{max} \sim 32$.
Therefore, our results indicate that enforcing causality 
helps to stabilize the calculation, and this may be even more important than using a good initial guess in this case.
However, using DMD is still advantageous, as it allows one to employ larger block time step sizes $dt_\mathrm{step}$. (Here, the block time step was growing exponentially, reaching $dt_\mathrm{step}=64$ in the last doubling.)
Minimizing 
the total computational cost of the calculation requires a careful choice of $dt_\mathrm{step}$, to avoid inflating the iteration number by using too many block time steps or risking instabilities and/or slow convergence (Fig.~\ref{fig:stability}(a)) by using too few.
Furthermore, we find that using $G_{0\mathbf{k}}$ as the ansatz often leads to larger bond dimensions in the intermediate iterations, slowing down the simulation. 
The self-consistency seems to go through some unphysical states with large $D$ before converging to the correct low-$D$ result.
This is possibly related to the discontinuity at the time point where the converged data are joined to the $G_{0\mathbf{k}}$ ansatz, and also might be related to the lack of damping in the latter.
While this problem can sometimes be cured by forcefully restricting $D$, using DMD bypasses the problem altogether, as one directly starts with an ansatz which is physically reasonable and does not introduce any significant discontinuities.

The correctness of our block time stepping solver is checked in Fig.~\ref{fig:stability}(b) by demonstrating energy conservation.
Note that there are no visible artifacts at the edges of the time steps, at the scale relevant for the energy variations. 

\begin{figure}[t]
	\centering
	\includegraphics[]{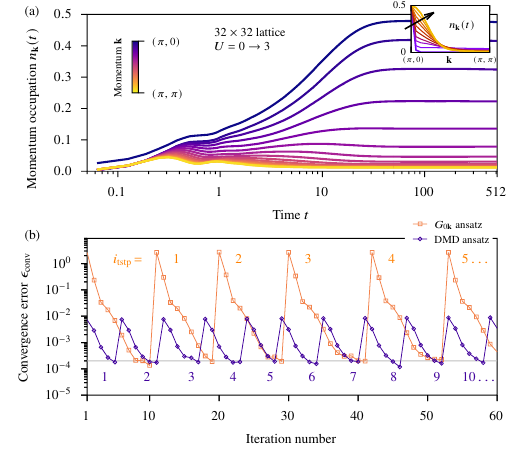}
	\caption{(a) Relaxation of momentum occupations $n_\mathbf{k}(t)$ after a $U=0 \to 3$ quench in a $32 \times 32$ Hubbard model up to $t_\mathrm{max}=512$ and for inverse temperature $\beta=100$. The momenta are taken along the direction $\mathbf{k}=(\pi,0)\to(\pi,\pi)$ (top to bottom). The inset shows snapshots of $n_\mathbf{k}(t)$ at $t \in \{0, 0.3, 1, 2, 4, 6.3, 9.4, 12.5, 18.8, 100, 400\}$. 
	(b) Convergence error $\epsilon_\mathrm{conv}$ for the few first block time steps within the last doubling ($t_\mathrm{max}=256\to 512$) of the simulation in (a). Presented are both the initialization of the extended domain with the noninteracting $G_{0\mathbf{k}}$ and with the DMD extrapolation. The numbers located next to the curves enumerate the block time steps, $i_\mathrm{tstp}$, with each point being a single self-consistency iteration. The gray horizontal line marks the convergence goal, which is $2 \times 10^{-4}$ in the shown last doubling ($1 \times 10^{-4}$ in the earlier doublings). 
	There are 5 doublings in total using block time step counts $n_\mathrm{step}=1,2,4,8,64$, corresponding to $dt_\mathrm{step}=16$ ($4$ in the last doubling), and starting from a global iteration for $t_\mathrm{max}=16$.}
	\label{fig:nkt_U3}
\end{figure}

The second key result is the ability to further extend $t_\mathrm{max}$ beyond the results presented in Ref.~\cite{Sroda2024} and doing so at larger interactions, $U>2$.
This is shown in Fig.~\ref{fig:nkt_U3}(a), where we plot the relaxation of the momentum occupations $n_\mathbf{k}(t) = -i G_\mathbf{k}^<(t,t)$ after a quench from $U=0$ to $U=3$.
We are able to reach an unprecedented simulation time, $t_\mathrm{max}=512$ inverse hoppings, with a high momentum resolution, and clearly resolve the system's thermalization.
We stress that we do not need to use slow interaction ramps, which were helpful to stabilize the global iterations in our initial implementation \cite{Sroda2024}, but became prohibitive for interactions above $U=2$ and for larger $t_\mathrm{max}$.
Unlike our approach in Fig.~\ref{fig:stability}, here, we find that it is best to keep the number of block time steps rather large (64 for the last doubling from 256 to 512) and the $dt_\mathrm{step}$ roughly constant across each doubling.
The rationale is to avoid the scenario of Fig.~\ref{fig:stability}(a), where convergence in each successive doubling is getting slower.
As we show in Fig.~\ref{fig:nkt_U3}(b), if $dt_\mathrm{step}$ is properly chosen, each block time step converges in a small and predictable number of iterations (about 5-10).
This is the behavior usually found in the conventional time-stepping scheme \cite{Nessi}, but note that here our $dt_\mathrm{step}=4$ is almost two orders of magnitude larger than what is used in the conventional implementation. 
A single iteration of our scheme is, however, more expensive in computational time (although cheaper in memory) than the conventional one, at least for modest $t_\mathrm{max}$. Still, above a certain $t_\mathrm{max}$, the QTT-NEGF wins over conventional schemes which either run out of memory or get similarly slow due to $\mathcal{O}(t_\mathrm{max}^3)$ scaling of the computational time.
The asymptotic scaling of QTT-NEGF is also expected to be better, with a potential saturation of the bond dimensions \cite{Sroda2024}. However, this is problem-dependent and requires further studies to reach a definitive conclusion.

Another clear advantage of the DMD ansatz compared to $G_{0\mathbf{k}}$ is shown in Fig.~\ref{fig:nkt_U3}(b).
Here, we plot the convergence error 
for the block time step calculations in
the last doubling (256 to 512). 
Starting a step with the DMD ansatz lands the convergence error immediately around $10^{-2}$, two orders of magnitude lower than if the calculation was started with $G_{0\mathbf{k}}$.
Correspondingly, for converging a block time step initialized with DMD one needs roughly half the iterations that would be needed when using $G_{0\mathbf{k}}$.
While one could also run new DMD after each converged step, to always use the newest information, we do not find this to be necessary in practice.
Our tests show that the DMD prediction is of similar accuracy on the whole doubled domain, as evidenced by Fig.~\ref{fig:nkt_U3}(b), which shows that the first iteration always stays at the error of $10^{-2}$. This error generically grows only very mildly up to the very last time step.

\begin{figure}[t]
	\centering
	\includegraphics[]{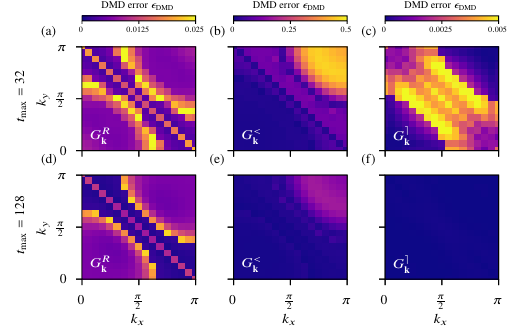}
	\caption{The normalized difference $\epsilon_\mathrm{DMD}(\mathbf{k}) = |G^\alpha_{\mathbf{k},\mathrm{DMD}}-G_{\mathbf{k},\mathrm{conv}}^{\alpha}|_\mathrm{F} / |G_{\mathbf{k},\mathrm{conv}}^{\alpha}|_\mathrm{F}$ between the DMD prediction $G^\alpha_{\mathbf{k},\mathrm{DMD}}$ and the final converged solution $G_{\mathbf{k},\mathrm{conv}}^{\alpha}$ across the Brillouin zone. Each column corresponds to a different Green's function component: the retarded ($\alpha=R$), lesser ($\alpha={<}$), and the left-mixing component ($\alpha={\rceil}$). The top row presents data for a prediction from $t_\mathrm{max}=16$ to $t_\mathrm{max}=32$ and the bottom from $t_\mathrm{max}=64$ to $t_\mathrm{max}=128$. The Green's function being extrapolated corresponds to the simulation of Fig.~\ref{fig:stability}.}
	\label{fig:dmdibz}
\end{figure}

\begin{figure}[t]
	\centering
	\includegraphics[]{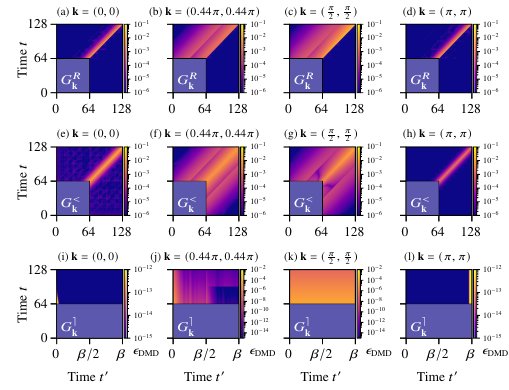}
	\caption{The absolute difference $\epsilon_\mathrm{DMD}(t,t') = |G^\alpha_{\mathbf{k},\mathrm{DMD}}(t,t')-G_{\mathbf{k},\mathrm{conv}}^{\alpha}(t,t')|$ between the DMD prediction $G^\alpha_{\mathbf{k},\mathrm{DMD}}$ and the final converged solution $G_{\mathbf{k},\mathrm{conv}}^{\alpha}$ in the full two-time plane at select $\mathbf{k}$ points. Each column corresponds to a different $\mathbf{k}$ point and the rows represent different real-time components: the retarded ($\alpha=R$), lesser ($\alpha={<}$), and the left-mixing component ($\alpha={\rceil}$). The data correspond to a prediction from $t_\mathrm{max}=64$ to $t_\mathrm{max}=128$, same as in bottom row of Fig.~\ref{fig:dmdibz}.}
	\label{fig:dmddiff}
\end{figure}

To finalize our analysis, we illustrate in Figs.~\ref{fig:dmdibz} and \ref{fig:dmddiff} how the DMD prediction differs from the final converged solution.
First, in Fig.~\ref{fig:dmdibz}, we plot the normalized Frobenius norm difference between the two as a colormap of the Brillouin zone.
Clearly, the DMD prediction is mostly accurate up to $10^{-2}$, with the left-mixing component having a lower error $\sim 10^{-3}$ and the unoccupied $\mathbf{k}$ points in the lesser component having a larger error, $\sim 10^{-1}$.
The former can be attributed to a simpler time dependence of the left-mixing component, which represents the decay of the initial equilibrium Matsubara-axis solution.
The latter is related to the way we extrapolate the unoccupied $\mathbf{k}$ points. 
Since for the lesser component, the diagonal values (occupations) can grow
 (Fig.~\ref{fig:nkt_U3}(a)), the DMD has a hard time describing this behavior, as it represents the dynamical system by a combination of exponents. A growing function requires an exponent with eigenvalues outside the unit disk which leads to an instability.
Hence, when extrapolating the lesser component for unoccupied momenta, we actually use a symmetry to transform it into $G_{\mathbf{k}}^>$ and extrapolate the latter, which is likely responsible for the disparity between occupied and unoccupied points.
As a proxy for identifying unoccupied $\mathbf{k}$ points, we use the relative norms of the greater and lesser components, defining a given $\mathbf{k}$ as unoccupied when $|G^>_\mathbf{k}|_\mathrm{F} > |G^<_\mathbf{k}|_\mathrm{F}$.
An alternative solution to extrapolate unoccupied $\mathbf{k}$ could be using the complete momentum information within DMD, as the full $\mathbf{k}$-dependent Green's function is bounded. To this end, one may add observable-function tensors corresponding to momenta to each operation in Fig.~\ref{fig:sketch}.

In any case, the error of the present scheme decreases as longer time data are supplied to the DMD, also for the unoccupied $G_{\mathbf{k}}^<$, as evidenced by comparing the two rows of Fig.~\ref{fig:dmdibz}.
Furthermore, the error is generically larger closer to the Fermi surface, as expected.
At the Fermi surface the DMD has less information, as the quasiparticle energies are closer to 0, leading to a slow decay. Instead, far away from the Fermi surface we always have more time information, and hence the prediction is better.
This is one reason which makes it difficult to apply the memory truncation techniques \cite{Schueler2018,Stahl2022} to lattice calculations. The truncation time close to the Fermi surface needs to be rather large, and gets larger with higher momentum resolution, as there are more points resolved close to the Fermi surface.
This is an argument in favor of compression techniques which are emerging as the preferred approach for lattice simulations.

Finally, in Fig.~\ref{fig:dmddiff} we show the absolute difference between the DMD ansatz and the converged solution on the full two-time plane at select $\mathbf{k}$ points for all real-time components.
The prediction accuracy is quite satisfactory, with the errors on the order of $\sim 10^{-2}$ being concentrated mostly along the diagonal for the $G^R$, $G^<$, i.e., at the points where the functions have appreciable value (far from the diagonal they decay to very small values, much below $\epsilon_\mathrm{conv}$).
Interestingly, the error does not grow appreciably when moving along the diagonal, consistent with the fixed error of the initial iteration observed in Fig.~\ref{fig:nkt_U3}(b).
The prediction for $G^\rceil_\mathbf{k}$ is clearly much more accurate, owing to the simpler structure of the mixed component.
While the accuracy of our predictor step could possibly be improved, for instance, by using a different algorithm, e.g., the Estimation of Signal Parameters via Rotational Invariance Techniques (ESPRIT) \cite{Erpenbeck2025}, or simply a different procedure than in our Fig.~\ref{fig:sketch}, we do not expect that it would significantly improve the convergence, which already is stable and requires few iterations per time step (Fig.~\ref{fig:nkt_U3}(b)).
Future improvements of the QTT-NEGF approach should rather focus on optimizing the solver to minimize the computational time spent on a single iteration. This is the main remaining bottleneck, since the memory cost is not posing an issue anymore.

\section{Conclusions}\label{conclusions}

We have developed a predictor-corrector scheme for quantics-tensor-train nonequilibrium Green's function (QTT-NEGF) simulations, which generates accurate initial predictor guesses using dynamic mode decomposition (DMD). 
Applying it to the time-dependent Hubbard model, we demonstrated its effectiveness in stabilizing and accelerating simulations of interacting lattice systems.
We achieved a stable propagation up to $t_\mathrm{max}=512$ inverse hoppings on a $32\times32$ lattice and at interaction strengths up to $U=4$.
This performance not only surpasses our previous implementation~\cite{Sroda2024} but also exceeds the limits of conventional matrix-based methods, establishing our approach as the current state of the art in NEGF calculations of large translationally invariant systems with full memory kernel.
Furthermore, to our knowledge, our work constitutes the first direct integration of quantics-tensor-train compression with DMD extrapolation, allowing one to use exponentially fine time resolution in the DMD and offering a promising route to denoise QTTs.
The presented methodology (Sec.~\ref{dmd}) is general and not restricted to our NEGF application.

Our predictor-corrector scheme is based on a \emph{block time stepping} strategy \cite{Inayoshi2025}, in which the Kadanoff-Baym equations (KBE) are solved on successive block time steps that are orders of magnitude larger than the individual time steps in matrix-based NEGF implementations. The scheme preserves the fine quantics resolution, and thus in fact treats a huge number of very fine time steps. The initial guess for a given block time step is produced by a DMD based extrapolation procedure. In practice, for a reasonable step size, the first iteration error is already $\mathcal{O}(10^{-2})$ in Frobenius norm across the Brillouin zone, and the convergence per block time step becomes predictable and efficient, with reduced iteration counts compared to initialization with the noninteracting Green's function $G_{0\mathbf{k}}$. Use of the DMD extrapolation also limits the transient growth of the bond dimensions during the self-consistent iterations. The improved initialization, combined with causality-enforcing block time stepping, enables doubling schemes with large $d t_\mathrm{step}$ and removes previous stability constraints. This shifts the focus of future optimizations entirely to the tensor-network operations.

Beyond algorithmic improvements, this work highlights that compression-based approaches are better suited for lattice simulations than memory truncation schemes. In the latter, slow quasiparticle dynamics near the Fermi surface sets a prohibitively large scale for the memory kernel.
The QTT-NEGF framework opens the door to simulations reaching unprecedented timescales at high momentum resolution,
enabling controlled explorations of long-time dynamics and nonequilibrium steady states in correlated lattice systems.
We expect that the QTT-NEGF approach, augmented by DMD prediction and block time stepping, will establish itself as a robust and scalable framework for diagrammatic simulations of lattice problems.

\begin{acknowledgments}
M.\'S.\ and P.W.\ acknowledge support from SNSF Grant No.\ 200021-196966.
K.I.\ was supported by JSPS KAKENHI Grant Nos. 23KJ0883 and 25K17307, Japan.
M.S.\ acknowledges support from SNSF Ambizione Grant No.\ PZ00P2-193527.
H.S.\ was supported by JSPS KAKENHI Grants No.~22KK0226, and No.~23H03817 as well as JST FOREST Grant No.~JPMJFR2232, Japan.
The calculations were performed on the beo05 cluster at the University of Fribourg.
The QTT-NEGF implementation is written in Julia \cite{bezanson2017julia} and is based on the ITensor library \cite{itensor} and libraries developed by the tensor4all group \cite{Fernandez2024,tensor4all}.
\end{acknowledgments}

\bibliography{ref}

\end{document}